\documentclass[12pt,preprint]{aastex}

\newcommand{\OI}{O\,{\sc i}}
\newcommand{\OIII}{O\,{\sc iii}}
\newcommand{\CII}{C\,{\sc ii}}
\newcommand{\CI}{C\,{\sc i}}
\newcommand{\SiII}{Si\,{\sc ii}}

\newcommand{\NII}{N\,{\sc ii}}
\newcommand{\NIII}{N\,{\sc iii}}

\slugcomment{}

\shorttitle{The far--IR spectrum of Sgr~B2}
\shortauthors{Goicoechea, Rodr\'{\i}guez-Fern\'andez \& Cernicharo}

\begin{document}

\title{The far--IR spectrum of Sagittarius B2 region:\altaffilmark{1}\\
Extended molecular absorption, photodissociation and photoionization}

\author{Javier R. Goicoechea\altaffilmark{2}}
       
\affil{Departamento de Astrof\'{\i}sica Molecular e Infrarroja,
      Instituto de Estructura de la Materia, CSIC, Serrano 121, 
         28006 Madrid, Spain} 

\email{javier@damir.iem.csic.es}

\and
\author{Nemesio J. Rodr\'{\i}guez-Fern\'andez\altaffilmark{3}}

\affil{Observatoire de Paris - LERMA. 61, Av. de l'Observatoire, 
75014 Paris, France}


\and
\author{Jos\'e Cernicharo\altaffilmark{2}}

\altaffiltext{1}{Based on observations with ISO, an ESA project with 
instruments funded by ESA Member States  (especially the PI 
countries: France, Germany, the Netherlands and the United Kingdom) 
and with participation of ISAS and NASA.}

\begin{abstract}

We present large scale 9$'$~$\times$~27$'$ ($\sim$25~pc~$\times$~70~pc) 
far--IR observations around Sgr~B2  using the 
\textit{Long--wavelength spectrometer} (LWS) on board the
\textit{Infrared Space Observatory} (ISO).
The spectra are dominated by the strong
continuum emission of dust, the widespread molecular absorption of light 
hydrides (OH, CH and H$_2$O) and  the fine structure lines of 
[\NII], [\NIII], [\OIII], [\CII] and [\OI].
The widespread dust emission is  
reproduced by a cold  component (T$_d$$\simeq$13--22 K) together with a
warm component (T$_d$$\simeq$24--38 K) representing $\lesssim$10 \% of the
dust opacity.
The fine structure line emission reveals a very extended component of 
ionized gas. The  [\OIII]52~$\mu$m$/$88~$\mu$m and
[\NIII]57~$\mu$m$/$[\NII]122~$\mu$m line intensity ratios 
show that ionized gas has an averaged  electron density of 
$\sim$240~cm$^{-3}$. The ionizing radiation
can be characterized by a
hard but diluted continuum, with effective temperatures of 
$\sim$36000~K and a Lyman continuum photon
flux of $\sim$10$^{50.4}$~s$^{-1}$.
The spatial distribution of the ionizing sources 
respect to the extended cloud and the clumpyness of the medium
determine the large scale effects of the radiation.
Photo--Dissociation Regions (PDRs) can
be numerous at the interface of the ionized and neutral gas.
The analysis of the [\CII]158~$\mu$m
and [\OI]63 and 145~$\mu$m lines 
indicates a  far--UV radiation field of G$_0$$\simeq$10$^{3-4}$
and a density of $n_H$=10$^{3-4}$~cm$^{-3}$ in these PDRs.
The widespread OH lines are reproduced by nonlocal radiative 
transfer models for clouds of moderate volume 
density ($n_{H_2}$$\simeq$10$^{3-4}$~cm$^{-3}$)  
at T$_k$$\gtrsim$40--100~K.
PDR models can explain the  enhanced 
column density of species such as
H$_2$O, OH and O$^0$.  However, they fail
to  reproduce the observed 
NH$_3$/NH$_2$/NH$\simeq$100/10/1 abundance ratios.
For N--bearing species it seems that shock chemistry has to be invoked.
The molecular richness in the outer layers of Sgr~B2 is  probed by the
ISO-LWS Fabry--Perot ($\sim$35~km~s$^{-1}$) detections towards Sgr~B2(M),
where more that 70 lines from 15 molecular and
atomic species are observed at high signal to noise ratio.

\end{abstract}


\section{Introduction}

\begin{figure}
\plotone{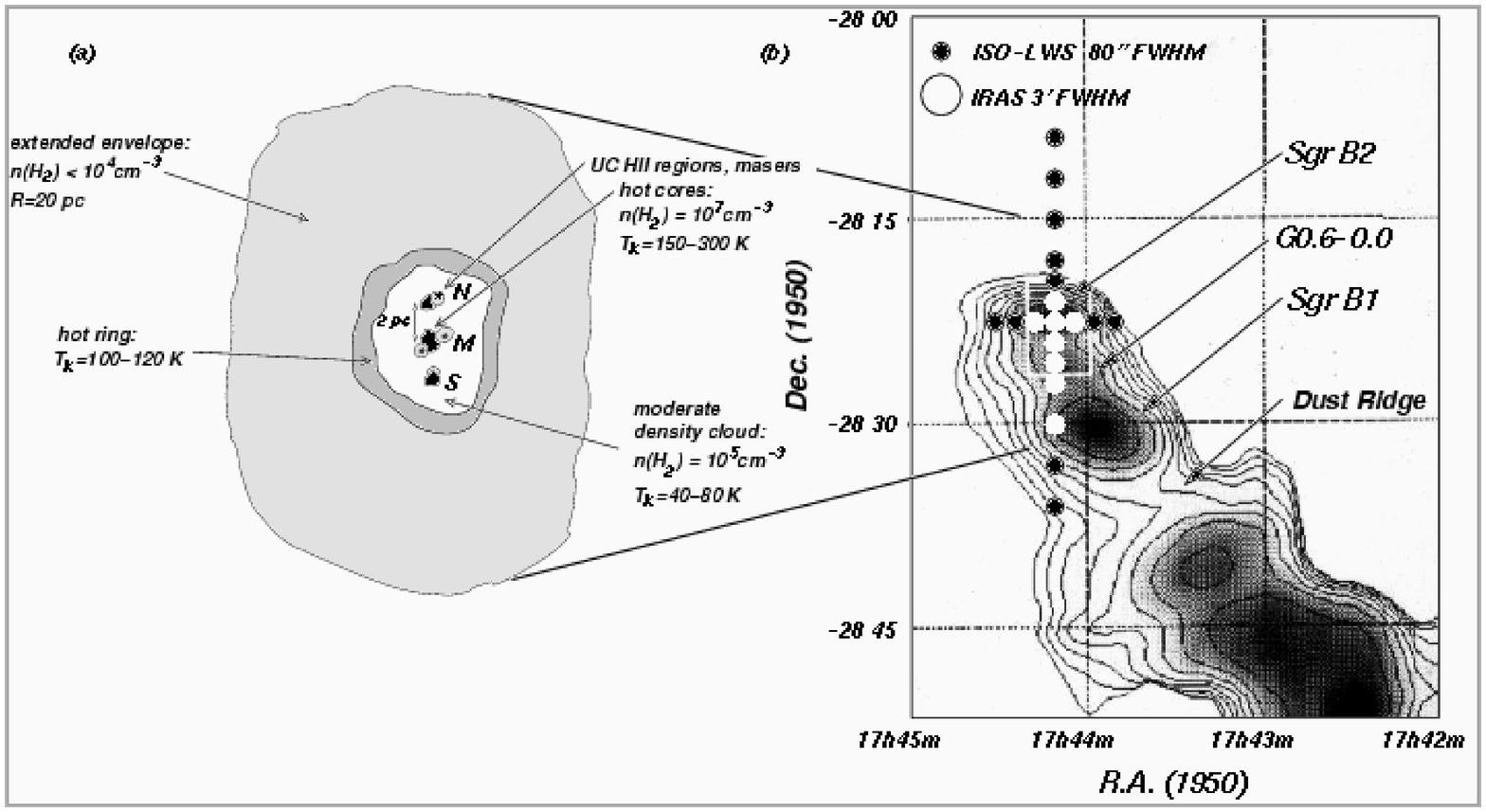}
\caption{ (a): Sketch showing the different structures and components in
the Sgr~B2 region. Hot cores are shown  black shaded and HII regions
are the structures enclosing the stars (adapted from H\"{u}ttemeister
et al. 1995).
(b): Large scale IRAS image at 60~$\mu$m (Gordon et al. 1993)
and ISO-LWS target positions across Sgr~B2 region. \label{sketch}}
\end{figure}

The Sagittarius~B (Sgr~B) complex is located in the inner 
400 parsec (pc) of the Galaxy, sometimes referred as the Central Molecular
Zone (Morris \& Serabyn 1996), at  $l\sim0.6^o\pm0.2^o$ galactic 
longitudes. It contains the well known 
Sgr~B2, B1 and G$0.6-0.0$ radio--sources
(see Fig.~\ref{sketch}b). Sgr~B1 lies to the south of the complex and
opposite to Sgr~B2, it is dominated by extended radio features
(see Mehringer et al. 1992). 
The G$0.6-0.0$ region is situated between Sgr~B2 and B1. The velocity
of the ionized gas in G0.6--0.0 is also intermediate between that
of Sgr~B1 and B2 suggesting that these regions are physically associated.
Large scale continuum studies show that Sgr~B2 is associated with the
brightest emission (Pierce--Price et al. 2000) and the most massive cloud
of the Galactic Center region (GC)
($10^7$~M$_{\odot}$, Lis \& Goldsmith 1990). 
In the following, we assume that it is situated at 
$\sim$100~pc from the dynamical center of the Galaxy for a distance of 
8.5~kpc (Kerr \& Lynden-Bell 1986). 

Figure \ref{sketch}a shows a schematic representation of the main
components within Sgr~B2. In the central region there are
three dust condensations labelled as Sgr~B2 north (N), middle (M) and 
south (S) which are situated in a imaginary North--to--South line of
$2'$($\sim$5 pc). 
They contain  all the tracers of on--going star formation: 
ultracompact HII regions created by the UV field of newly born $OB$ 
stars (cf. Benson \& Johnston 1984; Gaume \& Claussen 1990),
X--ray sources associated with  HII regions,  X--ray
sources with neither radio nor IR counterpart 
(Takagi et al. 2002),
hot cores (HC) of dense material (T$_k$=150--300~K; 
n$_{H_2}$$\simeq$10$^7$~cm$^{-3}$) and  embedded 
proto--stars (cf. Vogel et al. 1987; Lis et al. 1993), 
molecular  maser emission in  H$_2$O,  OH, H$_2$CO, CH$_3$OH and SiO 
(cf. Mehringer \& Menten 1997 and references therein) 
and high far--IR luminosity
($\geq$7$\cdot$10$^6$~L$_{\odot}$; Thronson \& Harper 1986). 
These components are embedded in a moderate density 
(n$_{H_2}$$\simeq$10$^{5-6}$~cm$^{-3}$) cloud
of $\sim$10~pc in size (Lis \& Goldsmith 1991;
H\"{u}ttemeister et  al. 1993).
The temperature in the moderate density cloud decreases with the distance
from
80 to 40~K except in a ring structure of warm gas (T$_k$=100--120~K)
with a radius of $\sim$4~pc 
(de Vicente et al. 1997).
These internal  regions  are surrounded by 
an extended lower density ($n_{H_2}$$\leq$10$^4$~cm$^{-3}$)
envelope ($\sim$15$'$), hereafter the Sgr~B2 envelope, 
of warm gas (T$_k$$\geq$100~K; H\"{u}ttemeister et al. 1995).

The origin of the observed rich chemistry in the 
Sgr~B2 envelope  and its heating mechanisms are far
from settled and several scenarios have been proposed.
Low--velocity shocks have been traditionally invoked to explain the
enhanced gas phase abundances of  molecular species 
such as SiO or NH$_3$ 
and the differences between gas and dust temperatures in the Sgr~B2 envelope
(Mart\'{\i}n--Pintado et al. 1997; Flower et al. 1995).
The origin of shocks in Sgr~B2  have been associated either 
with large scale  cloud--cloud collisions (Hasegawa et al. 1994)
or with small scale wind--blown bubbles produced by evolved massive stars
in the envelope itself (Mart\'{\i}n--Pintado et al. 1999).

The effect of the  radiation in the Sgr~B2 envelope has been traditionally
ruled out because of differences in the gas and dust temperature,
the unusual chemistry, and the absence of thermal radio-continuum and 
ionized gas outside the HII regions and hot cores within the central 
condensations.
The ISO observations presented here reveal the presence of an extended
component of ionized gas detected by its fine structure line emission.
The presence of widespread UV and X--ray fields illuminating 
large portions of Sgr~B2 could
trigger the formation of PDRs and XDRs in the
interface between the ionized gas and the self--shielded neutral layers
and could influence in the selective heating of the molecular gas.
The complexity of the region possibly allows a combination of
different scenarios and excitation mechanisms to coexist within the
whole complex.

In this paper we study the large--scale properties of  Sgr~B2 region 
and the effect of the far--UV radiation by analyzing its far--IR spectrum 
(43 to 197~$\mu$m). In Sect. 2, we summarize the ISO observations
and the data reduction. The resulting spectra are presented and analyzed
in the following sections: dust emission (Sect. 3), fine structure lines 
(Sect. 4) and molecular lines (Sect. 5).
Several techniques have been used in the analysis: gray--body fitting, 
photoionization models, molecular radiative transfer models 
and  comparisons with PDR models. A general overview 
and a brief summary is given in Sect. 6.

\section{Observations and Data Reduction}

The far--IR wavelength range covers the spectral signature of several 
interesting phenomena which are difficult to observe from ground--based
telescopes. These include; the fine structure lines of atoms and ions,
the high--$J$/fundamental  rotational  lines of heavy/light molecules,
the low energy bending modes of carbon clusters and the continuum 
emission peak for the  bulk of star forming  regions.
Therefore, we proposed to use the LWS spectrometer (Clegg et al. 1996)
on board ISO (Kessler et al. 1996) to study the large scale distribution and
of the dust, the ionized gas, the neutral gas and the molecular
content around the Sgr~B2 region.

The present study includes the observations of our open time programs
in different positions of the Sgr B complex. In addition we also
present some LWS 
observations retrieved from the
public ISO Data Archive\footnote{see http://www.iso.vilspa.esa.es/ida} (IDA)
observations.
The main observations are a $9'\times27'$ raster map  that 
targets 19 individual positions of the complex (see Fig.~\ref{sketch}b).
The cross-like LWS/grating  map is centered near Sgr~B2(M) position
at $\alpha= 17^h44^m10.61^s$, $\delta=-28^o22^{'}30.0^{''}$ [J1950].
Offsets between consecutive positions are  $90''$, excepting North--South
points with $|\Delta\delta|\geq450''$ for which a $180''$ spacing was
selected. 
The central position of our maps, Sgr~B2(M),
has been widely studied at higher resolution by different LWS/FP
observations (see references in Table 5).

\subsection{LWS AOT L01 Observations}

\begin{figure}
\centering
\includegraphics[angle=0, width=12cm]{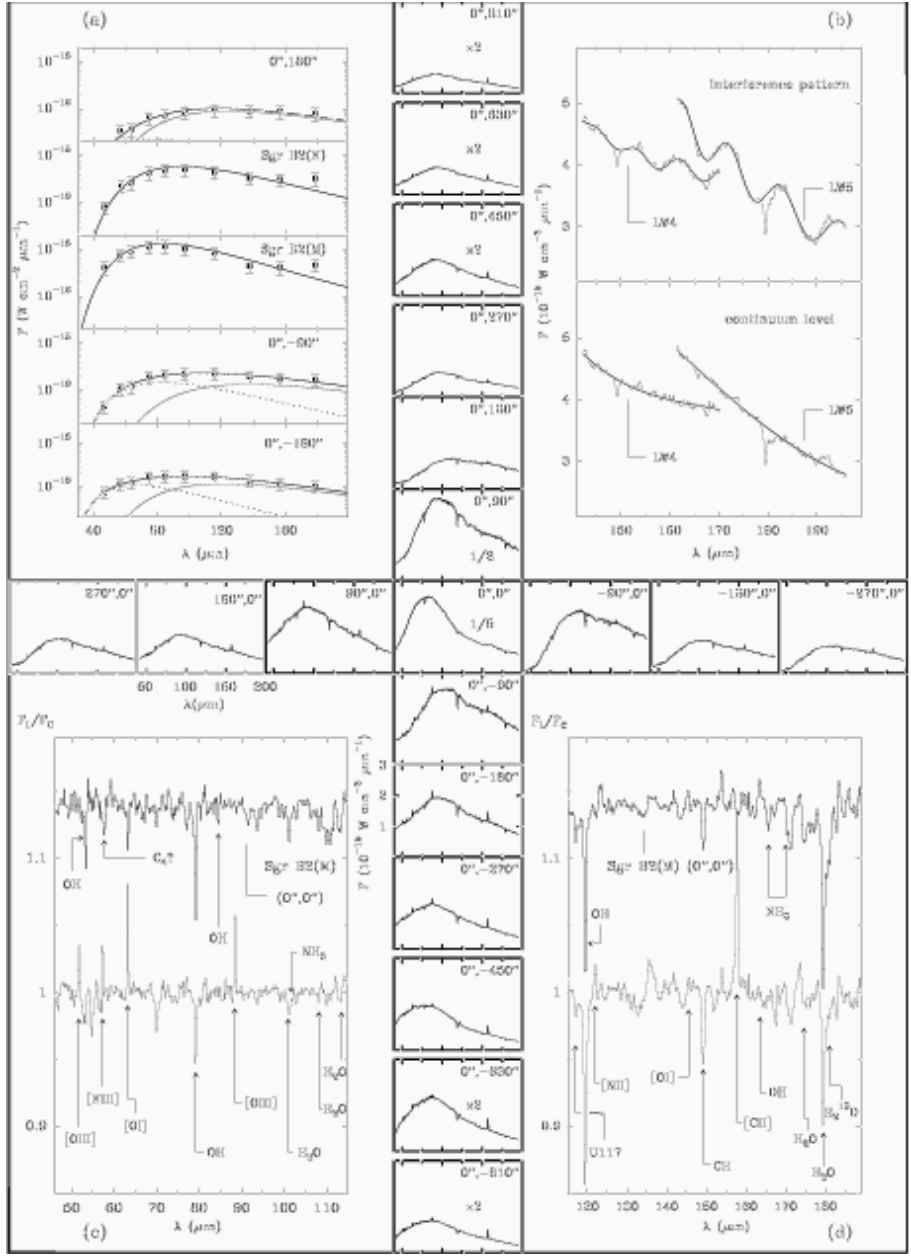}
\figcaption{\small  Raster map obtained with the LWS grating between
43--197~$\mu$m  with a spectral resolution of $\sim$1000~km~s$^{-1}$.
In each box, offset positions are given in arcsec
respect to the $(0'',0'')$ central position at:
$\alpha= 17^h44^m10.61^s$, $\delta=-28^o22^{'}30.0^{''}$ [J1950].The 
intensity scale corresponds to the flux (in units of 
10$^{-16}$~W~cm$^{-2}$~$\mu$m$^{-1}$) and the abscissa to the wavelength 
in $\mu$m.
(a) Averaged continuum flux of each LWS detector and gray-body best fits
(black) for some selected positions. Dotted line corresponds to the warm
component and continuous grey line to the cold component.
The error bars correspond to a 30\% of flux uncertainty.
(b) $Top$: Observed fringing in the long-wavelength detectors.
 $Bottom$: Continuum level after defringing the spectra.
(c) and (d) Comparison of LWS grating spectra of Sgr~B2(M) 
(0$''$,0$''$) [black] and an average of 
(90$''$,0$''$)+(0$''$,--90$''$)+(0$''$,--180$''$) adjacent positions
[grey]. A polynomial baseline have been removed.  \label{Fig_polvo}}
\end{figure}

The cross--like map was carried out during August 1996 and February 1997 
(target dedicated time numbers [TDTs] 28702130, 28702131, 46900233 and 
46900234) using the Astronomical Observation
Template (AOT) L01 with a spectral resolution of 0.29 $\mu$m for
the 43-93 $\mu$m  range (detectors SW1 to SW5) and 0.6 $\mu$m for 
the 80-197 $\mu$m range (detectors LW1 to LW5).
It uses all 10 LWS detectors with  beam sizes ($\Omega_{LWS}$)
around 80$''$. 
The typical flux accuracy varies from 10 to 50 \% 
depending on the source geometry, the source flux and the particular 
detector (Gry et al. 2002).
Based in the overlapping regions, the 
agreement in the flux measured by different detectors is found to be
better than 10\%.
Shifting factors $\leq10\%$ have been applied in some 
selected detectors to yield a smoothed and aligned continuum spectrum 
(Fig.~\ref{Fig_polvo}). 
Only detector LW2 showed intensities too high 
by $\sim20\%$ relative to the neighboring detectors and had to be scaled 
by a factor of 0.8. 
The spectra were oversampled at 1/4 of a resolution element. 
The long-wavelength data have been defringed for the interference
pattern systematically seen in the AOT L01  spectra of extended
sources or point sources which are offset from the optical axis 
(Swinyard et al. 1996, Fig.~\ref{Fig_polvo}b).

Depending on the position in the cross-raster, the far--IR spectrum 
exhibits several OH, CH and H$_2$O rotational lines and several
fine structure lines.
We have clearly detected  the  [\OI]63, 145~$\mu$m and [\CII]158~$\mu$m
lines in all positions, except in Sgr~B2(M). In addition, lines coming 
from higher excitation potential\footnote{Excitation potential of observed
ionic species are (in eV): 11.26(\CII), 14.53(\NII),
29.60(\NIII), 35.12(\OIII)} ions such as [\NII]122, 
[\NIII]57, [\OIII]52 and 88 $\mu$m lines are also detected. 
The OH ($\sim$79~$\mu$m), [\OI]63, [\NIII]57 and [\OIII]88~$\mu$m lines lie
in the overlapping regions of two different LWS detectors. As fluxes agreed
within $\sim$15~\%, we averaged both determinations. Line
fluxes were extracted by fitting gaussians after removing a polynomial 
baseline for each detector coverage.
Emission  line fluxes are listed in Table \ref{gr-fluxes}.

\subsection{LWS AOT L04 Observations}

The LWS/FP  instrument have been used to 
investigate the atomic and molecular features in Sgr~B2(M) at a 
higher spectral resolution ($\sim$35 km s$^{-1}$). The majority of detected lines
(see Fig.~\ref{survey-fp}) have been observed in our AOT L04 ISO proposals.
The TDTs are: 32201428, 32701751, 47600907, 47600908, 47600909, 47601001, 47601002,
and 46900332. However, an extensive inspection and reduction of about 
+50 TDTs\footnote{The additional AOT L04
observations analyzed were taken by ISO during orbits: 498, 494, 467, 
464, 462, 326, 327, 326, and 322 while the additional AOT L03 obsvervations
were taken during orbits: 849, 847, 845, 838, 836, 509, 508, 507, 506,
504 and 476.} from the public IDA
have been carried out in order to add some 
detections and average all available lines (see Polehampton 2002 for the 
AOT L03 detections).
The LWS/FP spectrum of Sgr~B2(M) includes the [\OI], [\CII] and [\OIII]  
lines (see Fig.~\ref{ionic}a and \ref{ionic}b) also detected at lower resolution in the 
LWS/gratings. In the case of the [\OIII]52 and 88 $\mu$m lines, its
clear LWS/FP identification confirms the doubtful detection made in the
gratings and shows the importance of increasing spectral resolution. 
A declination raster (TDT 32201429) of some 
molecular lines was also carried out up to $\Delta\delta=\pm270''$.
Fig.~\ref{Figuron-OH}a shows the raster in the
$^{16}$OH 119.442 $\mu$m line.
The richness of the LWS/FP spectrum of Sgr~B2(M) suggests that the far--IR
spectra of other observed sources can also be much richer than the
observed at the grating resolution.

\subsection{Data Reduction}

AOT L04  products have been processed and compared trough 
Off-Line-Processing (OLP) pipeline versions 6.0 to 10.1. There are no 
major differences except that recent pipelines produce $<10\%$ less 
absorption in some lines due to continuum level differences from one OLP 
to another. For the bulk of the observations, the  LWS/FP continuum flux 
in the wavelength coverage of each individual line deviates by $<$30\% 
and this has be taken as the FP flux calibration error. 
As recommended by the LWS handbook (Gry et al. 2002), we
first checked with the continuum level of the AOT L01 observations,
and then, a polynomial baseline was fitted to each spectra and adopted as
LWS/FP continuum level. 

AOT L01, L03 and L04 data were analyzed using the 
ISO Spectral Analysis Package 
 (ISAP). Typical routines include: deglitching spikes due to cosmic rays,
oversampling and averaging individual scans, defringing the
long-wavelength detector AOT L01 scans, removing baseline polynomials and
extracting line fluxes by fitting gaussians.

\section{Widespread far--IR continuum emission: results}

All positions around Sgr~B2(M) [(0$''$,0$''$)] present their continuum
emission peak between 90 and 100 $\mu$m (Fig.~\ref{Fig_polvo}), 
indicating that the bulk of observed dust has a relatively cold temperature. 
The  far--IR luminosity in the map 
is $L_{LWS}\simeq8.5\cdot10^6$ L$_{\odot}$.
The strongest IR  positions are Sgr~B2(M) and (N)[$\sim$(0$''$,90$''$)],
they contribute with $\sim$28\%  and 
$\sim$14\% to $L_{LWS}$, respectively. 
The rest of the positions show
decreasing continuum fluxes with increasing distance to Sgr~B2(M).
For the same distance to Sgr~B2(M), 
the southern points of the cloud have larger fluxes than the Northern
ones, while the dust emission in the East-West direction
is more symmetrical.

In order to estimate and better constrain the dust temperature and
the associated column density of material, we have modeled the 
observed continuum spectrum as a sum of two gray bodies.
A single gray body crudely fits the observed emission in any
position. The total continuum flux in the model is given by:
\begin{equation}
  S_\lambda=
(1-e^{-\tau_\lambda^{w}})\;B_\lambda(T_{w})\;\Omega_{w}
+ (1-e^{-\tau_\lambda^{c}})\;B_\lambda(T_{c})\;\Omega_{c}    
\label{eq-flujoBB}
\end{equation}  
where $c/w=i$ stands for cold/warm dust components, B$_\lambda$(T$_i$) is the 
Planck function at a temperature T$_i$, $\tau_\lambda^i$ is the 
continuum opacity and $\Omega_i$ is the solid angle subtended by the $i$
dust component.
We have expressed $\tau_\lambda$ at far--IR wavelengths as a function
of the 30 $\mu$m opacity using a  power--law with exponent $\beta$
($\tau_\lambda=\tau_{30} (30/\lambda)^\beta$).
In addition,  $\tau_{30}$ can be written as a function of the visual 
extinction ($A_V$) as $\tau_{30}=0.014\, A_V$ (Draine 1989).
Thus, $\tau_\lambda$ is given by:
\begin{equation}
   \tau_\lambda^{c/w} \; = \; 
0.014\;A_V\;\left(30/\lambda\right)^\beta
\label{eq-opacidad}
\end{equation}
Taking into account the large extension of the dust emission in the
region we have considered that both dust components fill the beam for
all the observed positions ($\Omega_i$=$\Omega_{LWS}$). Note that  
Eq. \ref{eq-flujoBB} applies for all positions but not for
Sgr~B2(M,N) that will be discussed below.


We have tried to fit the continuum emission with $\beta$ 
between 1.0 and 2.0, which 
are the expected emissivity exponents for silicates and graphite 
grains (Sptizer 1979).
We obtain satisfactory fits in this range of $\beta$ values.
However,  fits
obtained for $\beta$$\sim$1 are slightly better ($\chi^2$ two times lower).
In addition,  the visual extinction derived for $\beta$$\sim$2 are more
than a factor of 10  larger than the extinction expected from 
the molecular column densities. 
Thus, $\beta$$\gtrsim$2 seems unrealistic for far-IR wavelengths. 
$\beta$ values between  1--1.5 are in agreement with those derived from ISO
observations of other GC clouds by Lis \& Menten (1998). Note that 
$\beta$$\sim$2 has been obtained from the
optically thin emission at 350 and 800~$\mu$m around Sgr~B2 
(Lis \& Carlstrom 1994; Dowell et al. 1999).
However, the central regions of the cloud ($\lesssim180''$) are characterized by 
$\tau_{100}\gtrsim1$, so that the continuum is optically thick 
in most of the far--IR wavelengths. 
Hence, the dust emission observed by ISO
basically arises in the external layers of the cloud, i.e., the
extended envelope that veils the dense star forming regions, while 
a considerable fraction of the
submillimeter (submm) emission comes from the dense regions of Sgr B2.
In addition, the extended dust component observed by ISO is
partially filtered out by the  beam switching submm observations. 
Thus, far--IR and submm observations 
could trace different dust components.
In the following, all  calculations have been carried out  for  $\beta$
values in the range 1.0--1.5.

Table \ref{tab-ratios} lists the lower and upper limits to
the visual extinction.
The extinction varies from  A$_V$$\gtrsim250$~mag for the
positions within a radius of 90$''$ ($\sim$4~pc)
to A$_V$$\gtrsim50$~mag for  positions within 270$''$ ($\sim$10~pc).

Table \ref{tab-Tdust} gives the dust temperatures 
(unlike the visual extinction, T$_d$ is only weakly dependent on $\beta$).
The  spectral energy distributions are
best fitted  with a dust component with a temperature of 13--22~K and
a warmer component with a temperature of 24--38~K.
The higher dust temperatures are those measured in the southern regions.
The warmer component contributes less than 10~\%  to the total extinction. 
For comparison, Gordon et al. (1993) derived  
T$_d$$\simeq$19~K for a smaller region ($95''\times270''$) 
using millimeter observations and graybody analysis, 
while they obtained IRAS 100~$\mu$m/60~$\mu$m color temperatures 
of $\sim$35~K.
The IRAS observations are 
more sensitive to the GC diffuse dust (Gordon et al. 1993).
The properties of this diffuse component (temperature and opacity)
agree with those derived for the warm component in our fits to the far--IR
emission of Sgr~B2.

\subsection{Sgr~B2(M) and Sgr~B2(N)}

One of the main properties of the Sgr~B2 central region is the
tricky Sgr~B2~(N)$/$(M) continuum flux 
ratio ($N/M$) as a function of
the observed wavelength. The observational evidence
that $N/M<1$ at 53~$\mu$m (Harvey et al. 1977), while $N/M>1$ 
at  1300~$\mu$m (Goldsmith et al. 1987), can be explained if Sgr~B2(N)
is embedded behind the dust and gas envelope of Sgr~B2(M). 
Thus, the Sgr~B2(N) line of sight will have a larger column
density of dust producing a greater emission 
at millimeter wavelengths.
However, at far--IR wavelengths, part of the warm dust emission from
Sgr~B2(N) will be absorbed by the cooler foreground dust 
associated to Sgr~B2(M), resulting in 
a $N/M<1$ ratio (Thronson and Harper 1986; Goldsimth et al. 1990). 
From the present LWS observations of Sgr~B2(M,N),
we found $N/M=0.3$ at 57~$\mu$m and $N/M=0.7$ at 178~$\mu$m.
These ratios confirm  the importance of
dust opacity  even at $\sim$180~$\mu$m.
Due to the large dust opacity in the 
 obscured Sgr~B2(M,N) positions, we have only fitted a 
single graybody to extract the averaged dust temperature and opacity.
From the fits  we infer $\tau_{100}\simeq3.8\pm0.4$
(T$_d$$\simeq$31$\pm$1~K) and 
$\tau_{100}\simeq5.3\pm0.6$ (T$_d$$\simeq$26$\pm$1~K)
for Sgr~B2(M) and (N), respectively (see Fig.~\ref{Fig_polvo}a).

\section{Fine structure lines: results}

\subsection{Extinction corrections}

The large H$_2$ column densities (up to 10$^{25}$~cm$^{-2}$) found 
across Sgr~B2 region,
suggest that even in the far--IR, fine structure lines can suffer 
appreciable extinction. In addition, the averaged interstellar extinction
toward the GC, with 
A$_V$$\simeq$25~mag 
(cf. Schultheis et al. 1999), also contributes
to the attenuation of atomic emission.
In this work we estimate limits to the extinction in each
position by using two approximations.
A lower limit to the extinction can be obtained using the
 [\OIII]52  to [\OIII]88~$\mu$m (hereafter [\OIII]52/88) line intensity ratios. 
This ratio can not be lower than $\sim$0.55, which is the value
obtained in the low electron density limit if lines
are optically thin. 
For those positions where the [\OIII]52/88 ratios are lower than
the lower limit,
we derive a minimum visual extinction of 
$\sim$20~mag for very distant positions  ($\geq7.5'$) from Sgr~B2(M),
while a minimum extinction of $\sim$100~mag inside the $15'$ diameter cloud
have been found.

A more direct estimation of the prevailing extinction was obtained
from the continuum analysis (Sect. 3).
Note that the lower limit to A$_V$ derived from the [\OIII] 52/88 ratios
is consistent with that derived from the dust models 
(Table \ref{tab-ratios}).
In the subsequent discussion we have corrected the line 
intensities by the extinction limits 
presented in Table \ref{tab-ratios}.

\subsection{The ionized gas}

The [\OIII] lines shown in Fig.~\ref{ionic} (see also
Fig.~\ref{Figuron-OH} for the [\NIII] and [\NII] lines) reveal an extended
component of ionized gas in the
southern and eastern regions of Sgr~B2.
In particular, the [\OIII]88~$\mu$m emission extends $\sim$13.5$'$
($\sim$35~pc) 
to the south of Sgr~B2(M).
The smooth decrease of the [\OIII]88~$\mu$m intensity
as a function of the distance to Sgr~B2(M) suggest that the
major contribution to the observed flux arises in an extended 
component of ionized gas rather than in compact sources. 
We also note that the G0.6-0.0 and
Sgr~B1 radio sources could contribute to the ionization
in the southern positions.

To study the properties of the ionized  gas  
we have analyzed the [\OIII]52/88
and the  [\NIII]57 to [\NII]122~$\mu$m (hereafter [\NIII]/[\NII])
line intensity  ratios.
Table \ref{tab-ratios}  lists both ratios for all positions
after correcting for extinction.

\subsubsection{Electron densities}

We have used the [\OIII]52/88 ratio to estimate
the electron density in the observed  sources 
(see Rubin et al. 1994).
The [\OIII]52/88 ratio derived from our observation 
varies between $\sim$ 3 for the central sources to
$\sim$ 0.5 for positions located far from Sgr~B2(M).
Comparing these [\OIII] ratios with Fig. 1 of Rubin et al. one finds
electron densities ranging from $\sim$10$^3$~cm$^{-3}$ to $\sim$ 50~cm$^{-3}$ 
for the sources located close and far from Sgr~B2(M), respectively. 
The average electron density in all observed positions is 240~cm$^{-3}$.

For Sgr~B2(M) itself,
the [\OIII] lines are hardly detected with the LWS in grating mode.
Nevertheless, Fig.~\ref{ionic}a shows their unambiguous
LWS/FP  detections towards Sgr~B2(M).
Both [\OIII] lines appear centered at $v_{LSR}\simeq50\pm15$ km
s$^{-1}$ and, as it could be expected,
do not show emission/absorption at more negative velocities produced
by the foreground gas in the line of sight.
From the LWS/FP [\OIII] line intensities
and correcting the [\OIII] 52/88 line intensity ratio by the 
$\sim$1000~mag of visual extinction derived for Sgr~B2(M),
we found an electron density of $\sim$10$^{4.3 \pm 1.3}$~cm$^{-3}$.
As expected,  the densest ionized material seen in the far--IR
is located in the central star forming regions of Sgr~B2.

Mehringer et al. (1993), from 20 cm
interferometric observations ($26''\times15''$ in resolution),
detected a $\sim7'$ ($\sim$20 pc) halo of diffuse 
emission around Sgr B. The [\OIII]88 $\mu$m line
emission in Fig.~\ref{ionic} spreads   beyond the radio 
recombination lines contours
of Mehringer et al.

\subsubsection{Radiation temperatures}
\label{sub-rad}

We have used the [\NIII] 57 and [\NII] 122  $\mu$m  line intensities
and followed  the method described in Rubin et al. (1994) for 
ionization-bounded nebulae to derive the  effective temperature
of the ionizing radiation (T$_{eff}$). 
For each position,  we have determined the volume emissivities
of both the [\NIII]57 and [\NII]122 $\mu$m  lines for
the electron densities derived  from the [\OIII] 52/88 line
intensity ratio.
With these emissivities it is possible to derive the actual N$^{++}$ to 
N$^{+}$ abundance ratio.
From the analysis of Rubin et al. (see their Fig. 4) and from
the [\NIII]57 and [\NII]122 $\mu$m line intensities, it is
possible to derive T$_{eff}$.  Table \ref{tab-ratios} gives the derived
values at several positions. The largest T$_{eff}$ is obtained
towards the central positions ($\sim 36000$~K).

It has been pointed out by Shields \& Ferland (1994), that  
the fine-structure lines ratios observed in the GC can be reproduced
with higher T$_{eff}$ and a lower incident flux of ionizing photons
(low ionization parameters).
This would be the case if the ionizing radiation is diluted, 
i.e.,  if the medium is clumpy and inhomogeneous, and/or
the ionizing sources are located far from the ionized nebulae. 
Indeed, this is the situation found in the 
Radio-Arc region, also in the GC, by  Rodr\'{\i}guez-Fern\'andez et
al. (2001), where  an extended (40~pc~$\times$~40~pc) gas component
is ionized by hot ($\sim 35000$~K) diluted
radiation arising from the Quintuplet and the Arches clusters.
They concluded that the radiation 
reaches large distances due to the inhomogeneity of the medium.
If this applies also to the Sgr~B2 envelope, then, the T$_{eff}$ 
derived above should be considered as lower limits.

In this work it is not possible to perform a detailed study of the
geometry of the region due to the limited angular and spectral
resolution of the ISO data. Instead, we have performed some
simple photo-ionization model calculations  using
the MPE IDL CLOUDY Environment (MICE), which was developed by
H. Spoon at the Max-Planck-Institut f\"ur extraterrestische Physik 
(MPE)\footnote{MICE, SWS and the ISO Spectrometer Data Center at MPE 
are supported by DLR (DARA) under grants 50QI86108 and 50QI94023.}
and uses  CLOUDY 94 (Ferland 1996).
To define the shape of the continuum illuminating the nebulae we have taken
the stellar atmospheres modeled by Schaerer \& de Koter (1997).
First, we have modeled the Sgr~B2 cloud as an sphere with  the  
density law  as taken from Lis \& Goldsmith (1990):
\begin{equation}
7\times 10^{5}\;\;(cm^{-3})  \;\;\; 0.3 <R\;(pc)< 1.25  \\
\end{equation}
\begin{equation}
7\times 10^{5}\; \left(\frac{1.25}{R}\right)^2 + 2000 \;\;(cm^{-3})  \;\;\;
1.25<R\;(pc)<22.5 
\end{equation}
where $R$ is the radius in pc.
We have considered an ionizing source in the center of this sphere emitting
a Lyman continuum photon flux, $Q$(H), of 10$^{50.3}$~s$^{-1}$ 
(approximately equal to the total $Q$(H) in  
Sgr~B2(M,N), see Gaume et al. 1995).
To define the shape of the ionizing continuum
we have used a 37800~K atmosphere (similar to 
the maximum T$_{eff}$ derived with the nitrogen lines ratio).
The results show that the radius of the [\NII] region would not be larger
than 1~pc.
The same is true for models with 
$Q$(H)=10$^{51.3}$~s$^{-1}$ and a 41700~K atmosphere.
With such a dense and homogeneous model it is not possible to explain the
large extension of the ionized gas component observed with ISO. 
However, if the cloud around the newly born $OB$ stars 
is inhomogeneous and clumpy enough, the UV radiation field can
illuminate several surfaces along the line of sight
(Tauber \& Goldsmith 1990). This scenario was required in the past 
to explain  the first extended [\CII] and [\CI] observations 
in molecular clouds (e.g., Phillips \& Huggins 1981).
Moreover, the poor correlation between the HC$_3$N and C$^{18}$O
extended emission found around Sgr~B2 suggests that both the
envelope and the dense regions are clumpy and/or 
fragmented (Lis \& Goldsmith 1991; Goldsmith et al. 1992).

Instead of doing a complex three-dimensional inhomogeneous density
model, we have treated the different observed positions as independent
nebulae located at a distance from Sgr~B2(M) equal to their projected
distance in the plane of the sky.
The model assumes that the medium is inhomogeneous and that the radiation
arising from Sgr~B2(N,M) can reach all the observed positions but, of course,
it takes into account the dilution caused by the distance of the different
sources to Sgr~B2(M).
This is done by means of a ionization parameter, $U$, defined as:
$U=Q(H)/4\pi n_e c D^2$, where $n_e$ is the electron density, $c$
is the speed of light and D the distance of the nebula to the ionizing source.
We take $n_e$=240~cm$^{-3}$ (the average
density derived from the [\OIII] lines) for all positions.

Figure \ref{fig_4} illustrates the results of some of these
models in terms of the [\NIII]/[\NII] ratio versus the
distance in arcsec to Sgr~B2(M) for the north-south raster.
Black lines  show that the observed line ratios
can be reproduced with $Q(H) \simeq 10^{50.3}$~s$^{-1}$ arising from
the center of the cloud and T$_{eff}$ between 35500 (dashed lines) and 36300~K 
(dot-dashed lines).
Unfortunately,
the large error-bars due to the extinction uncertainties make difficult
a more precise analysis.
However it seems that the [\NIII]/[\NII] ratios in the
southernmost positions are somewhat higher than  would be expected
from the points closer to Sgr~B2(M) and  from model calculations.
This should be an effect of the presence of additional ionizing sources in
the Sgr~B1 area.
We have tried to model the effect of Sgr~B1 assuming that the effective
temperature of the radiation arising from these sources is similar to that 
arising from Sgr~B2(M) and to estimate a total ionization
parameter defined as:
\begin{equation}
U=\frac{1}{4\pi n_e c} \left(\frac{Q(H)_2}{D_2^2} +
\frac{Q(H)_1}{D_1^2}\right)
\end{equation}
where $Q(H)_2$ and $Q(H)_1$ are the Lyman continuum photons arising 
from Sgr~B2(M) and Sgr~B1, respectively, and $D_2$ and $D_1$ are
the distance of the observed sources to Sgr~B2(M) and Sgr~B1, respectively.
Grey lines in Fig.~\ref{fig_4} show the results of some of these
    combined models with $Q(H)_1=10^{49}$ or $10^{49.5} \;s^{-1}$
(Mehringer et al. 1992).
Including the effect of Sgr~B1 helps to explain the observed ratios in the
(0$''$,--450$''$) and (0$''$,--630$''$) positions, but the measured ratio in
(0$''$,--810$''$) is still  higher than expected.
Hence, additional ionizing sources in the southern region of Sgr~B2
cannot be ruled out.

In any case, it is important to remark that even taking into account 
the simplicity of the model and the dust extinction uncertainties,  
the agreement of the model with the observations is fairly good.
All  measured ratios lie in between the 35500 and the 36300~K  curves
(see Fig.~\ref{fig_4}).
We conclude that the whole Sgr~B complex is permeated by hot radiation 
arising mainly from Sgr~B2(M,N). 
There is also a contribution to the large scale ionization 
from sources located in the vicinity of Sgr~B1. 
Minor contribution from additional ionizing sources cannot be ruled out.
The long-range effects of the ionizing radiation can only be
understood if the medium is clumpy and inhomogeneous, but also if the
location and geometry distribution of the ionizing sources 
is appropriate to ionize predominantly the southern and eastern regions.

\subsection{Photodissociation regions}

The detection of a widespread component of ionized gas
suggests that numerous PDRs can exist in the interface between this ionized
material and the molecular gas throughout the cloud.
Even more, the prevailing far--UV radiation 
field and the X--ray emission 
can also be important in the heating and in the chemistry
of the neutral gas.
In the following sections we  analyze the fine structure
emission related with the PDRs and the molecular content of Sgr~B2
as seen by far--IR spectroscopy.

\subsection{The [\NII] vs. [\CII] correlation}

The chemistry and the heating of a PDR are 
basically controlled by the hydrogen gas density ($n_H$) and by the far--UV 
($6\; eV<h\nu<13.6\; eV$) radiation field. 
The main coolants in a PDR are the far--IR continuum emission of dust
and the [\CII] and [\OI] fine structure lines. Thus, 
their relative intensities can be used as a diagnostic of the  PDR conditions.
The first step, hence, is to find the way to distinguish the diffuse
ionized gas from the PDR one. \\

If ionized and neutral/PDR phases exist and/or are associated, 
[\CII] emission  can arise from both components (Heiles 1994).
For positions with [\NII] detections, the [\CII] emission coming
from the diffuse gas should scale with the [\NII] lines
because the bulk of N$^{++}$ emission arises in the 
low--density ionized gas ($\sim$75~\% according to Malhotra et al. 2001).
Figure~\ref{nii_cii} shows the good correlation
found in the Sgr~B2 region.
The lack of  [\CII] emission in the grating spectra of Sgr B2(M,N) is 
a combined effect of extinction, self--absorption in foreground clouds
and absorption of the continuum by  C$^{+}$ in low--excitation  
diffuse clouds (see the absorption in Fig.~\ref{ionic}a).
Those observations were not included in the correlation.
However, the line is well detected further from the central
position where the extinction is less considerable and there is
less background continuum to be absorbed by the diffuse component.
The resulting observational correlation is:
\begin{equation}  
   I(C^+)_{-11} \; \simeq \; 
5.2\cdot I(N^+)_{-11}+6.4 
\label{eq-correlation}
\end{equation}
where $I(C^+)_{-11}$ and $I(N^+)_{-11}$ are the 158 and 122~$\mu$m line
intensities in units of 10$^{-11}$ W~cm$^{-2}$~sr$^{-1}$.
A crude approximation to the [\CII] emission
arising in PDR gas can be estimated by assuming that
the second term in relation~\ref{eq-correlation} 
represent the averaged [\CII] emission in the PDRs.
For comparison, Malhotra et al. (2001) derived the theoretical scaling 
$I(C^+)= 4.3\cdot I(N^+)+I(C^+)_{PDR}$. However,
a linear correction for the diffuse [\CII] emission and the same
value for all the PDRs in the complex could be a rather crude approximation
to the structure and physical conditions of these PDRs.

\subsection{PDR diagnostics}

Once we have estimated the amount of [\CII] arising from PDRs,
we can compare the usual far--IR diagnostics: 
[\CII] 158, [\OI]63 and 145~$\mu$m lines and continuum emission,
with  theoretical PDR models to estimate the gas density and the
far--UV incident field.

\begin{figure}
\centering
\includegraphics[angle=0, width=12cm]{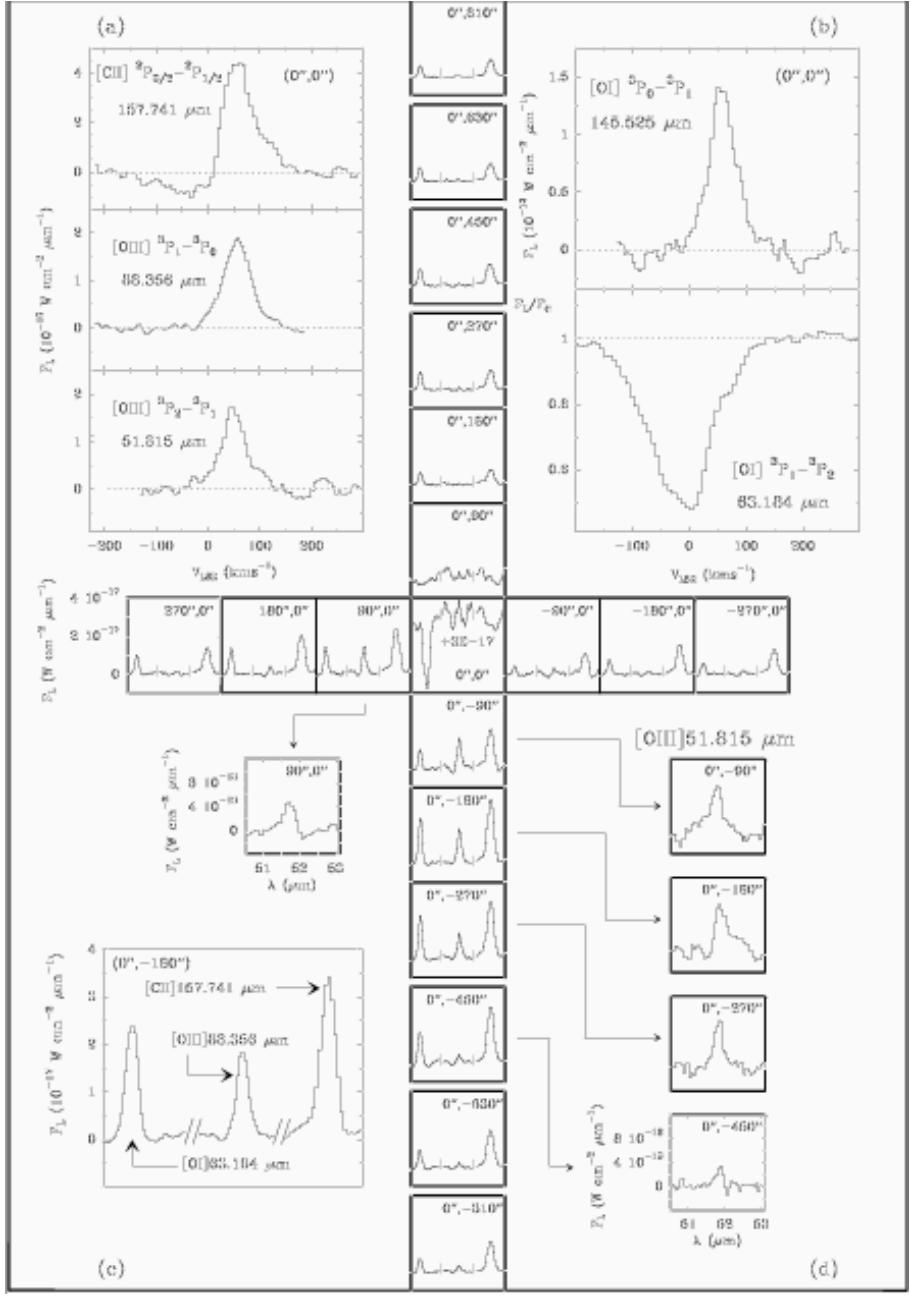}
\figcaption{\small Raster map obtained with the LWS grating of the
[\OI]63, [\OIII]88 and [\CII]158 $\mu$m lines. In each box, offset positions
are given in arcsec respect to the $(0'',0'')$ central position at:
$\alpha= 17^h44^m10.61^s$, $\delta=-28^o22^{'}30.0^{''}$ [J1950].The intensity scale corresponds to line flux and the abscissa to the wavelength. 
 (a) [\CII] and  [\OIII] lines detected with LWS-FP in $(0'',0'')$
 (b) [\OI] lines detected with LWS-FP in $(0'',0'')$.
 (c) Main features of the grating raster map labelled. 
 (d)  Raster map positions with clearest [\OIII]52 $\mu$m detections.  
\label{ionic}}
\end{figure}

\begin{figure}
\centering
\includegraphics[angle=-90, width=13cm]{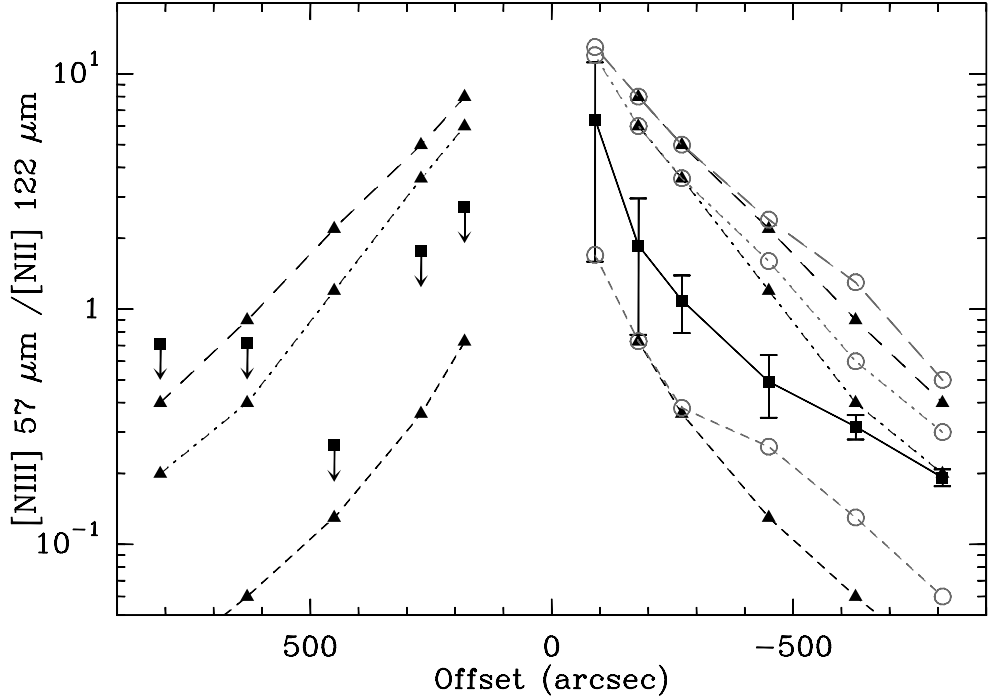}
\figcaption{Solid squares show the [\NIII]57~$\mu$m /[\NII]122~ $\mu$m 
lines ratios derived from the observations for the
north-south raster.
The arrows indicate that those points are  upper limits.
The error-bars are due to
the uncertainties in the extinction correction.
Upper limits for Sgr~B2(M) [(0$''$,0$''$)] and Sgr~B2(N) 
[(0$''$,90$''$)] are not significant
and therefore they are not shown.
The solid triangles and empty circles 
show the photo-ionization models results for 
$n_e$=240~cm$^{-3}$ taking into account Sgr~B2 (triangles) and
both Sgr~B2 and Sgr~B1 (circles).
Points from a same model are connected by lines. 
The black dashed line and the triangles 
show the model results for  T$_{eff}$=35500~K and 
$Q$(H)$_{2}$=10$^{50.3}$~s$^{-1}$.
The grey dashed line and circles represents
the results considering additional $Q$(H)$_{1}$=10$^{49.5}$~s$^{-1}$ 
from Sgr~B1.
The dot-dashed lines and the triangles 
represents the  results for $T_\mathrm{eff}$=36300~K
and $Q$(H)$_{2}$=10$^{50.3}$~s$^{-1}$. The grey line  with circles has
additional $Q$(H)$_{1}$=10$^{49.0}$~s$^{-1}$.
The long-dashed lines with triangles 
represents the  results for $T_\mathrm{eff}$=36300~K
and $Q$(H)$_{2}$=$10^{50.5}$~s$^{-1}$. The grey long-dashed line with 
circles has additional $Q$(H)$_{1}$=10$^{49.0}$~s$^{-1}$. \label{fig_4}}
\end{figure}

\begin{figure}
\centering
\includegraphics[angle=0, width=13cm]{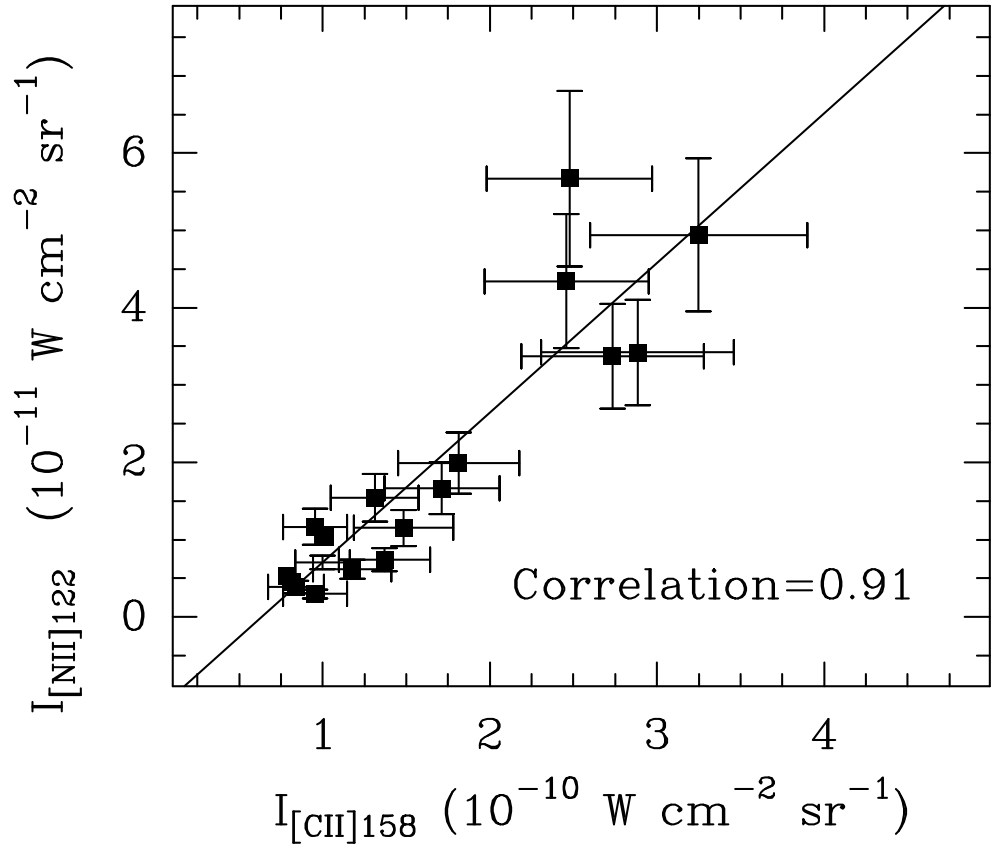}
\caption{Correlation between the [\NII]122~$\mu$m and [\CII]158~$\mu$m
lines. Each source position have been corrected for its minimum visual
extinction derived from far--IR continuum fits (see table \ref{tab-ratios}).
Sgr~B2(M,N) positions are not included (see text). \label{nii_cii}}
\end{figure}

\begin{figure}
\centering
\includegraphics[angle=0, width=13cm]{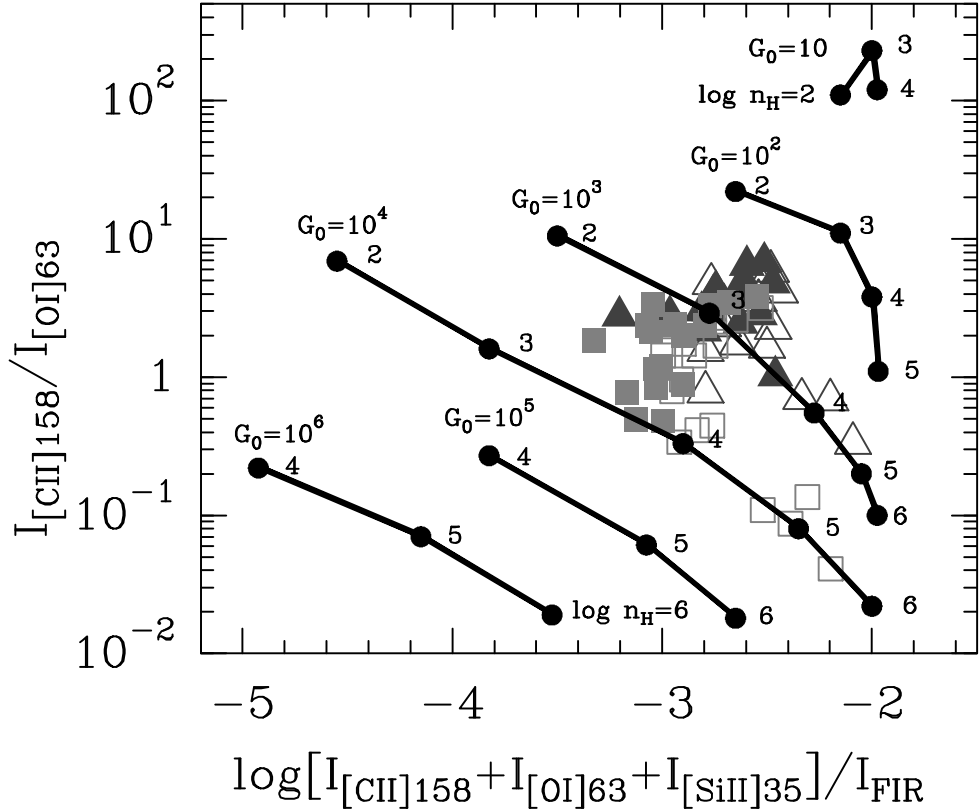}
\caption{[\CII]/[\OI] intensity ratio vs. ([\CII]+[\OI]+[\SiII])/far--IR 
for several PDR models of varying far--UV fields and hydrogen densities 
(from WHT90). The different points of Sgr~B2 are shown in gray. 
Dark gray triangles for calculations that consider all the observed [\CII] 
intensity at  each position and light gray squares for a mean [\CII] intensity
from PDRs of $6.4\times10^{-11}$~W~cm$^{-2}$~sr$^{-1}$ in all positions.
The different points represent the intensity ratios corrected
for the minimum (filled) and maximum (not filled) visual extinction 
(see Table \ref{tab-ratios}). \label{cii_oi_fir}}
\end{figure}

Because of the prominent foreground absorption in the [\CII] 158 and 
[\OI]63~$\mu$m lines observed towards Sgr~B2(M),
we have omitted\footnote{The LWS/FP spectra
of the [\OI]63~$\mu$m line at  $\Delta\delta=\pm180''$ away from Sgr~B2(M) 
shows the line in emission at Sgr~B2 velocities ($\sim$60~km~s$^{-1}$) 
and no foreground absorption (Lis et al. 2001). Here we 
 consider that outside the bright Sgr~B2(M,N) lines of sight, the 
bulk of observed [\OI] emission arises in Sgr~B2.} 
the (M,N) positions in the following discussion (see Vastel et al. 2002).
Fig.~\ref{cii_oi_fir} shows PDR models
predictions of the [\CII]/[\OI] ratio vs. [\CII]+[\OI]+[\SiII]/far-IR ratio
in terms of the density, $n_H$, and the far--UV incident flux,
(G$_0$; in units of the local interstellar value)
taken from Wolfire, Tielens and 
Hollenbach (1990, hereafter WTH90). Nevertheless, Simpson et al. (1997)
underlined that Si is highly depleted in WTH90 models so that
the effect of the not observed [\SiII] line across the region can be 
neglected.
Experimental points in Fig.~\ref{cii_oi_fir}
do not include the [\SiII]35~$\mu$m line 
because only the central position has been
observed (Goicoechea \& Cernicharo 2002, hereafter GC02). 
The uncorrected line
intensity ratio in Sgr~B2(M) is [\CII]158$/$[\SiII]35$\simeq$10, with 
the [\CII] intensity coming from the emission component (see Fig.~\ref{ionic}a
for the LWS/FP line).

Squares and triangles in Fig.~\ref{cii_oi_fir} show the parameter
space occupied by Sgr~B2.
Dark gray triangles represent line ratios that consider all the observed
[\CII] intensity at each position and light gray squares only consider
the mean PDR [\CII] emission derived from the correlation with [\NII].
The different points show the intensity ratios corrected
for the minimum (filled) and maximum (not filled) visual extinction 
(see Table \ref{tab-ratios}). The data scatter over the
G$_0$$\simeq$10$^{3-4}$ and $n_H$$\simeq$10$^{3-4}$~cm$^{-3}$ curves.
The expected PDR surface temperature for those values is $\simeq$300~K
(WTH90),
and it can reach $\sim$500~K if heating by photoelectrons ejected
from PAHs and very small grains is included in the  
models (Kaufman et al. 1999).

The possible errors in the PDR  parameters are 
influenced by the A$_V$ uncertainty of each line of sight, the estimation of
the  PDR  [\CII] emission  and  
optical depth effects such as the [\CII] absorption seen in 
Fig.~\ref{ionic}a. 
Additional uncertainty 
is due to the [\OI]63 $\mu$m line  intensity used as a PDR diagnostic
in Fig.~\ref{cii_oi_fir}. The line intensity can be larger  
if cold foreground  gas absorbs part of the emission
associated with Sgr~B2 also outside (M,N) lines of sight. The 
line could be also weaker (relative to the [\OI]145~$\mu$m
line) if the emission is saturated due to a high optical depth. 
Finally, the line  
intensity could also be  overestimated if the majority of the emission
arises from other inner  regions (mainly molecular) of the cloud and not
from the PDRs. In such a case, and assuming that the [\OI]63~$\mu$m 
line emission is very thick across Sgr~B2, it is difficult to estimate the
different contributions to the measured line intensity.
Therefore,
the main properties of the widespread PDRs in Sgr~B2
derived from Fig.~\ref{cii_oi_fir} should be
considered to cover the range of densities $n_H$=$10^{3-4}$~cm$^{-3}$
the far--UV incident flux $G_0$=10$^{3-4}$.\\

The ionized  component shown in Fig.~\ref{ionic}
proves the presence of an extended far--UV radiation field. Although
the PDR models can explain the fine structure line  intensities,
we have also considered the role of shocks in the excitation
of  [\OI] and [\CII] lines. Such shocks are known to be present
in Sgr~B2 (H\"{u}ttemeister et al. 1995; Mart\'{\i}n--Pintado et al. 1997).

The ratio of the integrated line emission ([\OI]63+[\CII]158) to
the integrated far--IR emission in PDRs can not be larger
than $\sim$10$^{-2}$ due to the low efficiency of the photoelectric 
heating mechanism. 
On the other hand, this ratio is at least an order of magnitude larger
if the lines arise in shocked gas 
(Hollenbach \& McKee 1989). Thus, 
the observational ratios in Fig.~\ref{cii_oi_fir} show that the 
extended [\CII] and [\OI] emission is dominated by the PDR scenario.

The absolute intensities 
can also be compared with PDR and shock models in the range
$n_H$=10$^{3-4}$~cm$^{-3}$.
The intensity of the [\CII]158 $\mu$m line associated
with PDRs estimated from the [\NII] vs. [\CII] correlation
is $\sim$6.4$\times$10$^{-11}$~W~cm$^{-2}$~sr$^{-1}$, which we
found to be consistent with the PDR model predictions of 
Hollenbach et al. (1991) for G$_0$$\sim$10$^3$.
Only $J$--shocks models (Hollenbach \& McKee 1989) predict some
[\CII] emission but it is orders of magnitude weaker than the observed
unless high shock velocities ($v_S$$\sim$100~km~s$^{-1}$) occur.
These velocities are not inferred from the line widths derived
from large--scale molecular observations at radio
wavelengths (see  H\"uttemeister et al. 1993 for NH$_3$ lines).  
Also the averaged absolute intensity of the [\OI]145~$\mu$m line,
$\sim$2.3$\times$10$^{-11}$~W~cm$^{-2}$~sr$^{-1}$ (Sgr~B2(M,N)
not included), agrees within a factor $\sim$2 with the PDR model
predictions. Finally, the low--velocity ($v_S$$\sim$10~km s$^{-1}$)
$C$--shocks models 
of Draine et al. (1983) predict weaker [\OI]145~$\mu$m line 
intensities. Comparing these two model predictions 
with the observed  intensity 
we estimate that the shock contribution to the [\OI] emission
in Sgr~B2 is 10--30~\%.

\section{The molecular gas}

Most pure rotational lines of  light  molecules  appear in the far--IR
and submm domains. The search for these species provides a crucial
insight on the   chemical pathways leading to the observed 
richness in molecular clouds. Besides, the far--IR absorption
measurements allow to trace more extended and lower density gas 
(i.e., the envelopes) than
the one observed in collision--excited emission surveys. The majority of
line surveys in Sgr~B2 have been concentrated in 
Sgr~B2 (M,N) and in the millimeter domain
(Cummins, et al. 1986; Sutton et al. 1991; Nummelin et al. 1998),
but less or nothing is known about the possible extended distribution of 
the light molecular species.

\subsection{Sgr~B2(M)}

The spectral resolution of the grating observations
is rather limited ($\sim$1000~km~s$^{-1}$) and produce strong
dilution in the search of  molecular features in most ISM sources.
In order to have an idea of the line density and molecular carriers
we have performed a search with the LWS/FP ($\sim$35~km~s$^{-1}$)
in Sgr~B2(M).
Fig.~\ref{survey-fp}  shows the most abundant species that can be
detected with ISO in the far--IR and gives insights of which could
be detected with the grating in other positions.
Tables \ref{survey1} and  \ref{survey2} 
list the observed transitions and their corresponding references
in the literature.
Molecular features include: Several \textbf{rotational lines}
of light O--bearing molecules such as H$_2$O, H$_2^{18}$O, OH, $^{18}$OH,
and H$_3$O$^+$, N--bearing molecules such as NH, NH$_2$ and NH$_3$ and other
diatomic species such as CH, HD or HF; Low energy
\textbf{bending modes} of non-polar carbon chains such as
C$_3$ or C$_4$ (only in the grating). The list of detected  molecules
could still increase  as  several weak 
features remain unidentified (see Polehampton 2002). 
Atomic features include the \textbf{fine structure lines} of 
[\OI], [\OIII] and [\CII].
In the next section we analyze the main molecular lines 
producing widespread absorption in  LWS/grating 
maps at lower resolution: OH, CH and H$_2$O.

\subsection{Widespread OH, H$_2$O and CH absorption}

Although the spectral resolution in the grating is limited,
the broad linewidths observed toward the GC 
($\gtrsim$30 km s$^{-1}$), and  Fig.~\ref{survey-fp} provided here,
could help in the detection of the most abundant
molecular species at large scale.
In Sgr~B2 we have confidently detected the prominent
absorption from  the
ground--state rotational lines of OH ($\sim$79 and $\sim$119~$\mu$m),
CH ($\sim$149~$\mu$m), H$_2$O ($\sim$179~$\mu$m) and H$_2^{18}$O
($\sim$181~$\mu$m; Cernicharo et al. 1997; 
possibly contaminated by H$_3$O$^+$, see
Goicoechea \& Cernicharo 2001)  
in almost all positions
(see Figs. \ref{Fig_polvo}c and \ref{Fig_polvo}d).
This implies that light hydrides are
present in a region as large as 25~pc~$\times$~70~pc.
Other molecular features produce less extended absorption/emission and their
definitive assignation will require better spectral resolution.
This second group of molecular lines include: 
H$_2$O ($\sim$180, $\sim$175, $\sim$108, and $\sim$101~$\mu$m); 
NH$_3$($\sim$170 and $\sim$166~$\mu$m); OH ($\sim$163, $\sim$84 and
$\sim$53~$\mu$m) and C$_4$ or C$_4$H ($\sim$58~$\mu$m;
Cernicharo, Goicoechea \& Benilan 2002).
The rest of the absorption feautures not present in Sgr~B2(M)
should be considered with caution.
Note that no far--IR rotational line of CO was successfully
assigned in any position at the signal-to-noise (S/N) ratio of the 
grating spectra.
However, we  have detected the CO $J$=7--6 line towards Sgr~B2(M) 
with an emission peak at $\simeq$+50~km~s$^{-1}$ (Goicoechea et al. 2003).
This and the LWS/FP H$_2$O lines from the warm molecular gas in front
of Sgr~B2(M) will be discussed in a forthcoming paper
(Cernicharo et al. 2004, in prep.).

Another striking result of the cross--raster map is the presence
of an unidentified absorption feature at $\sim$117~$\mu$m (U117) in all
positions near the OH 119~$\mu$m line 
(see Fig.~\ref{Figuron-OH}). This feature has a nearly constant 
absorption depth of  $\simeq$4$\%$ independent of the distance to the 
central position and is present in all individual scans. Prompted by 
this detection we inspected the public IDA$^2$  observations taken with the
LWS/FP in this wavelength range. A tentative carrier 
is presented in Sect. 5.3.

\subsubsection{Large scale OH absorption}

According to the chemical models, the hydroxyl radical, OH, is an important
intermediary in the formation of many molecules present in
both the dense n$_{H_2}$$\sim$10$^4$~cm$^{-3}$ (Bergin et al. 1995)
and the diffuse n$_{H_2}$$\sim$10$^2$~cm$^{-3}$ gas
(Van Dishoeck \& Black 1986). The typical OH abundance in dense
molecular clouds  is (0.1-1)$\cdot$10$^{-7}$. Enhanced abundances are 
predicted in molecular regions under $C$--shock activity (Draine et al. 1983) 
and in the outer layers of PDRs where H$_2$O is being 
photodissociated (Sternberg \& Dalgarno~1995). When applied to 
Sgr~B2(M), the $C$--shock models (Flower et al.~1995) can not reproduce the 
large OH/H$_2$O abundance ratio found in the warm envelope, which seems more
consistent with the PDR scenario (GC02).
The less self--shielded PDR layers will be also a source
of [\OI] by means of the OH photodissociation. Although in PDRs [\OI] 
can exist deeper in the cloud than OH  
(Sternberg \& Dalgarno 1995), a correlation between the 
emission/absorption of both species is expected in the outer layers
of the cloud and could be used
to follow the ionized--PDR--molecular gas relation.

\begin{figure}
\centering
\includegraphics[angle=0, width=14cm]{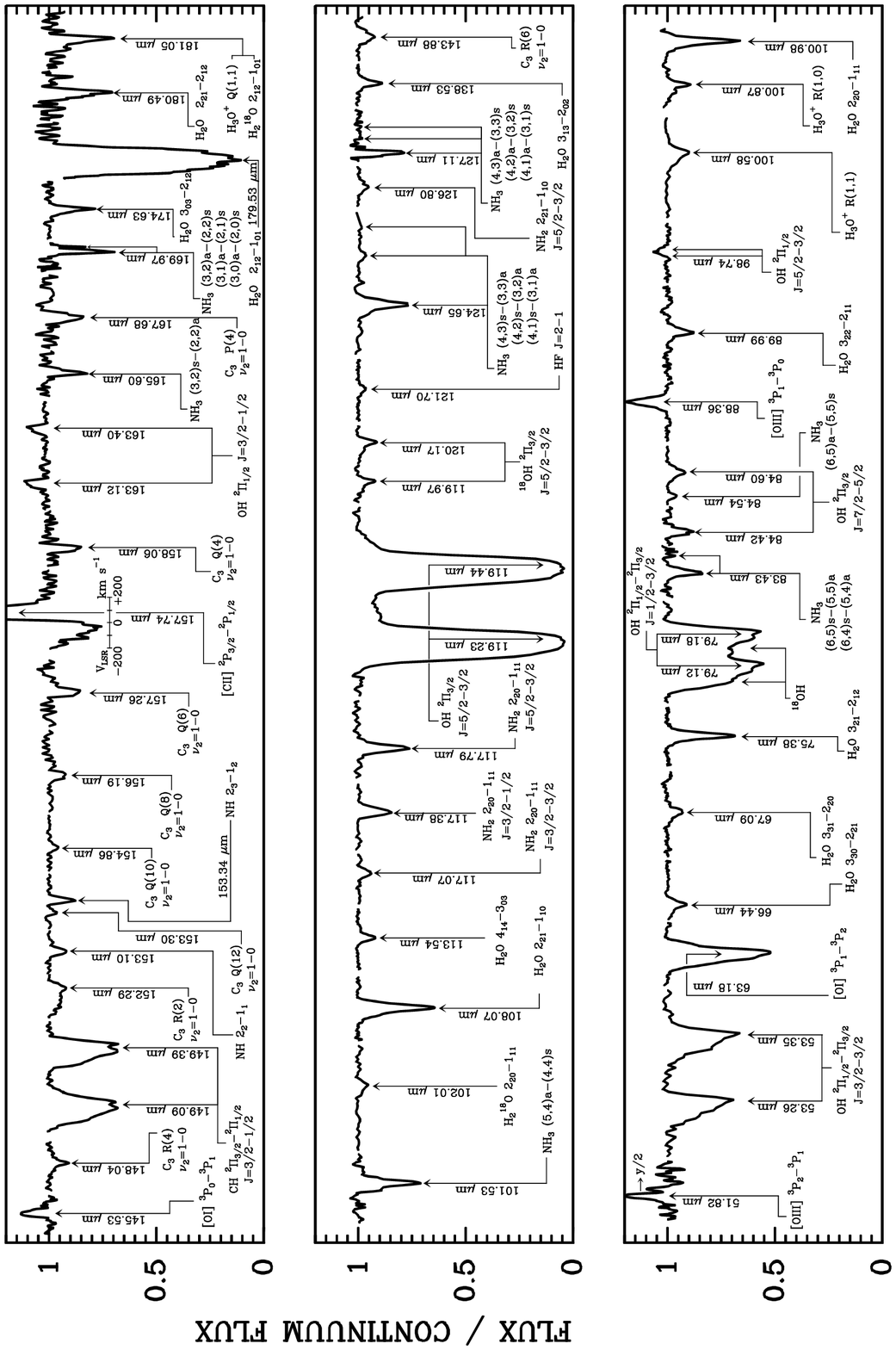}
\figcaption{\small Far--IR features in the spectrum of Sgr~B2(M) taken with the
ISO/LWS Fabry-Perot with a spectral resolution of $\sim$35~km~s$^{-1}$. 
Note the wavelength discontinuity of the spectrum after each line.
Some weak features remain unidentified (not shown) while some detections
have been discussed in other  works (see tables \ref{survey1} and 
\ref{survey2} and text references).\label{survey-fp}}
\end{figure}

\begin{figure}
\centering
\includegraphics[angle=0, width=8cm]{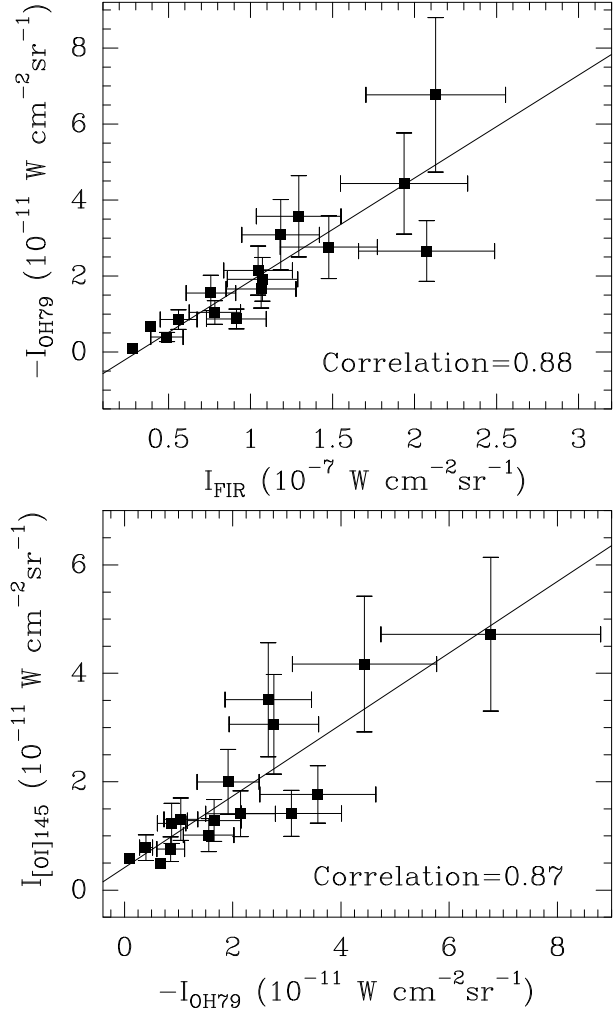}
\figcaption{$Top$: Observed correlation between the OH $\sim$79~$\mu$m
absorption and the far--IR continuum emission.
$Bottom$: Observed correlation between the [\OI]145~$\mu$m emission and
the OH $\sim$79~$\mu$m absorption.
Both figures exclude the (M,N) and $\Delta\delta=-90''$
obscured positions. \label{oh_corr}}
\end{figure}

The present cross--like grating maps show the OH 
$^2\Pi_{3/2}$ $J= 5/2 \leftarrow 3/2$  ground--state line
at $\sim$119~$\mu$m (Fig.~\ref{Figuron-OH}) and the 
$^2\Pi_{1/2} \leftarrow ^2\Pi_{3/2}$ $J= 1/2 \leftarrow 3/2$  
cross--ladder line at $\sim$79~$\mu$m (Fig.~\ref{varias_molec}) over  
the $\sim$25~pc~$\times$~70~pc region. Both lines may be blended with 
their $^{18}$OH and $^{17}$OH isotopomers due to the low resolution of the
spectra. The individual OH $\Lambda$--doubling line components can not 
be distinguished (only possible at the FP resolution, see the OH lines
in Fig.~\ref{survey-fp}).
The $\sim$119 $\mu$m line is detected in absorption in all directions
and remains strong through the map. The mean
line absorption depth $\simeq$20$\%$ is consistent with saturated  OH
lines.
On the other hand, the $\sim$79~$\mu$m absorption is observed below the
northern $\Delta\delta=630''$ position but its intensity is
 more sensitive to the position in the map (see
Fig.~\ref{oh_corr} and Fig.~\ref{varias_molec}). In fact, the $\sim$79~$\mu$m
line absorption depth seems to be tracing the dust continuum variations
through the region, but not necessary OH abundances, so that OH
column densities are correlated with far--IR emission
(upper box in Fig.~\ref{oh_corr}).
The absorption produced by the OH 
$^2\Pi_{1/2} \leftarrow ^2\Pi_{3/2}$ $J= 3/2 \leftarrow 3/2$
 line at $\sim$53 $\mu$m is  only clearly detected toward
Sgr~B2(M,N), absorbing $\simeq$3$\%$ of the dust continuum emission.
The low S/N  ratio of the spectra makes the 
identification rather difficult in  other positions.

Finally, an emission line centered at $\sim$163~$\mu$m have been found in 
almost all positions (see Fig.~\ref{varias_molec}). The line intensity,
2$\%$ of the continuum, appears to be 
constant across the region. The emission line probably arises from the 
OH $^2\Pi_{1/2}$ $J= 3/2 \rightarrow 1/2$ line which has been clearly 
detected in Sgr~B2(M) with the FP.
Hence, it is plausible that a fluorescence mechanism 
(absorption of photons in the $\sim$53~$\mu$m  line and
emission in the $^2\Pi_{1/2}$ line at $\sim$163~$\mu$m) similar
to that found in Sgr B2(M) by GC02, also  operates in the 
whole region favored by the low density of Sgr~B2 envelope 
and by the large far--IR continuum emission.

At the resolution of the grating  mode it is difficult to
discriminate which components along the line of sight are producing the 
absorptions [specially in the  lines coming from rotational ground--states
of light hydrides such as  OH, CH or H$_2$O (see the wide profiles
in Figs. \ref{varias_molec}b and d)]. For these lines, a 
considerable fraction of the absorption is produced by the low excitation 
clouds  in the  line of sight (see Neufeld et al. 2000).   
The $\Delta$$\delta$ = $\pm$270$''$ FP declination raster of the
OH 119.442~$\mu$m line (Fig.~\ref{Figuron-OH}a)
reveals that in all observed positions the absorption is dominated
by the cold foreground gas at negative velocities which is  not associated
with Sgr~B2. 
The line almost absorbs completely the   continuum emission and covers 
a broad velocity range, $\Delta$$v_{FWHM}$ $\simeq$~200~km~s$^{-1}$,
much larger than the FP resolution. Lines are thus saturated and imply 
high opacities. 

Another FP line declination raster was carried out in the OH $^2\Pi_{3/2}$
$J=7/2 \leftarrow 5/2$ excited state at 84.597~$\mu$m 
in order to follow up the excitation of the warm OH gas.
The line was  confidently detected  only toward  Sgr~B2(M,N).
This time, both lines are strictly centered at Sgr~B2 velocities
without appreciable foreground absorption.\\


The interpretation of the molecular absorption 
in the Sgr~B2 envelope is not obvious and realistic 
radiative transfer models  taking into account both  dust
and  molecular emission are needed.
The dust emission play a significant role because photons
emitted by dust grains can excite the far--IR rotational transitions
of molecules such as OH. In Sgr~B2(M,N) the dust grains can also
absorb the  photons emitted by the molecules.
For prominent  clouds like Sgr~B2, where the far--IR continuum 
emission and opacity are  substantial, these conditions mean that the 
external envelope can absorb both the dust emission and the molecular 
line emission  from the inner regions of the cloud
(if any emission escapes the core).

The present observations have been modeled with the same
nonlocal radiative transfer code (see Gonz\'alez-Alfonso \& Cernicharo
1993) that we used in the analysis of the OH lines towards Sgr~B2(M)
at higher resolution (GC02). 
As few OH rotational lines are clearly detected in the grating spectra,
we tried to estimate the approximate physical conditions leading
to the observed OH extended emission/absorption. 
The excitation temperature in the cross--ladder and $^2\Pi_{3/2}$ lines 
have to be lower than the  dust temperature (Table 2)
in order to see the lines in absorption.
In addition, the $\sim$163~$\mu$m emission
can not be very prominent while  the $\sim$84~$\mu$m 
line must be almost insignificant outside Sgr~B2(M,N) positions
(at the limited grating resolution). From this information we
found that the widespread OH component has a moderate density, 
$n$(H$_2$)=10$^{3-4}$~cm$^{-3}$, with lower limit temperatures in the range
100~K [Sgr~B2(M,N)] to 40~K (extended envelope). For a given 
temperature, larger densities will give
asymmetrical profiles in the $\Lambda$--doubling lines 
and emission lines will be also apparent in the cross-ladder transitions
(not observed at the FP resolution, Fig.~\ref{survey-fp}).  

These calculations  show that the PDR diagnostics 
and the OH nonlocal models yield similar physical conditions for the
outer layers of Sgr~B2.

The opacity in the cross--ladder absorption lines
is lower than in the intra--ladder transitions and absorption
depth differences from one position to other could  reflect OH column density 
variations together with dust emission differences 
(see upper box in Fig.~\ref{oh_corr}).
The observed lines can be fitted with total OH column densities
in the line of sight
in the range (5-10)$\cdot$10$^{15}$~cm$^{-2}$ outside the 
$\sim$15$'$ diameter cloud to (2--5)$\cdot$10$^{16}$~cm$^{-2}$ 
inside the $\sim$15$'$ cloud.

In addition, a correlation between the OH$\sim$79~$\mu$m 
absorption and the [\OI]145~$\mu$m line emission has been found in the region
(lower box in Fig.~\ref{oh_corr}).
The lower energy level of the [\OI] $^3$P$_0$--$^3$P$_1$ transition is 
at 230 K and
the emission is optically thin in most of the conditions. The LWS/FP
observations of the [\OI]145~$\mu$m 
line towards Sgr~B2(M)(Fig.~\ref{ionic}b)  show that the line emission
appears only at Sgr~B2 velocities ($v_{LSR}$$>$0~km~s$^{-1}$). Therefore,
most of the  [\OI]145~$\mu$m line emission arises from the warm
outer layers of Sgr~B2, where photodissociation is taking place.
On the other hand,  PDR models suggest that 
OH is preferentially produced in the external regions of the PDR.
The good  [\OI]145~$\mu$m vs. OH$\sim$79~$\mu$m correlation 
confirms that a large fraction of the warm [\OI] is 
produced by the   photodissociation of OH  and/or
it arises in the same warm OH layers.

\subsubsection{Large scale H$_2$O and CH absorption}

\begin{figure}
\centering
\includegraphics[angle=0, width=12cm]{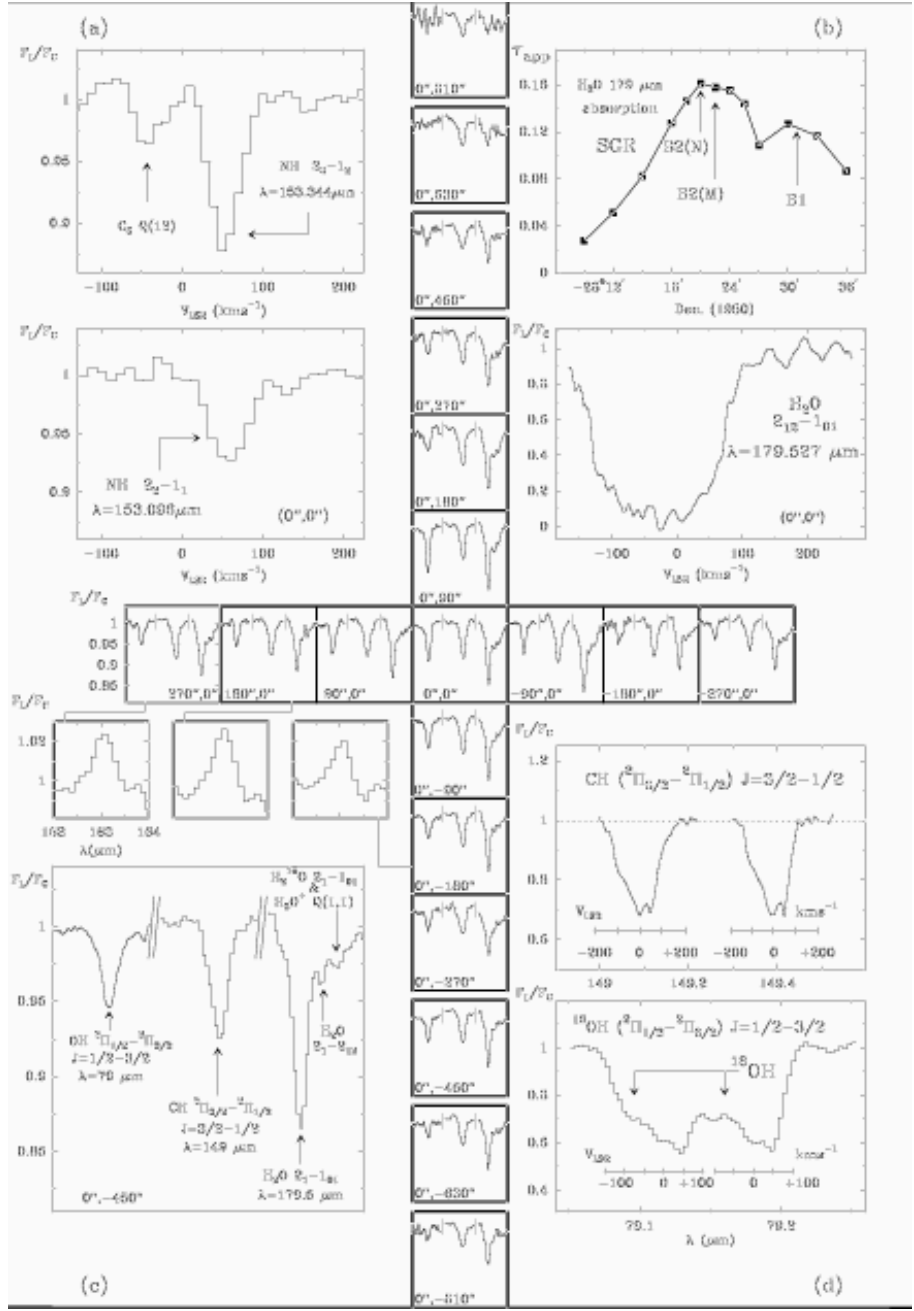}
\figcaption{\small Raster map obtained with the LWS grating of the
OH($\sim$79~$\mu$m), CH($\sim$149~$\mu$m) and H$_2$O($\sim$179~$\mu$m) lines.
 Detections of an emission feature at $\sim$163~$\mu$m are also presented.
In each box, offset positions are given in arcsec
respect to the $(0'',0'')$ central position at:
$\alpha= 17^h44^m10.61^s$, $\delta=-28^o22^{'}30.0^{''}$ [J1950].
The intensity scale corresponds to the continuum normalized flux and the
abscissa to the wavelength. 
(a) NH lines detected with the LWS-FP in $(0'',0'')$ around $\sim$153~$\mu$m.
(b) $Top$: H$_2$O 2$_{12}$-1$_{01}$ 
apparent opacity in the north--to--south declination raster with main 
sources position labelled.
$Bottom$: Fundamental o-H$_2^{16}$O line detected with the LWS-FP in
$(0'',0'')$ around $\sim$179~$\mu$m. 
(c) Main features of the grating raster map labelled. 
(d) CH and OH $\Lambda$--doublets detected with the LWS-FP in
$(0'',0'')$ around $\sim$149 and $\sim$79~$\mu$m, respectively.
\label{varias_molec}}
\end{figure}

Besides the OH absorption discussed in the previous
section, the fundamental lines coming from the rotational ground--states
of CH at $\sim$149~$\mu$m and ortho-H$_2$O 
at $\sim$179~$\mu$m (see also Cernicharo et al. 1997) are
also detected in the cross--like spectrum (Fig.~\ref{varias_molec}). 
Therefore, in addition to the widespread ionized gas and dust emission, 
the Sgr~B2 envelope can be characterized by its extended molecular content.

The H$_2$O $\sim$179~$\mu$m map represents further evidence 
that water vapor is extended in molecular clouds 
(cf. Cernicharo et al. 1994, 1997; Snell et al. 2000, Neufeld et al. 2003). 
In addition, the FP
observations of the  $\sim$179~$\mu$m line 
(Cernicharo et al. 1997; lower panel in Fig.~\ref{varias_molec}b) showed 
that water  is present in Sgr~B2, but also in the clouds of the line of 
sight (see also Neufeld et al. 2000 for the $\sim$557 GHz line). 
The full H$_2$O ($\sim$179~$\mu$m) map shows a smooth 
variation in the north--to--south direction suggesting that the 
H$_2$O content and/or the source excitation also change smoothly
across the different lines of sight. The H$_2$O absorption depth is
larger in the central positions around Sgr~B2(M) and near Sgr~B1
(see Fig.~\ref{varias_molec}b).\\

The last molecular line detected in all observed positions
is the CH $^2\Pi_{3/2}$$\leftarrow$$^2\Pi_{1/2}$ 
$J=3/2\leftarrow1/2$ line  at $\sim$149~$\mu$m.
The CH J=1/2 ground state of the $^2\Pi_{1/2}$
ladder produced by spin--orbit interaction has parity
characteristics of a rotational $^2\Pi_{3/2}$ level (Lien 1984).
Therefore, the  CH $\sim$149~$\mu$m line is the analogous to the
fundamental OH $^2\Pi_{3/2}$ $J= 5/2 \leftarrow 3/2$ intra--ladder line
at $\sim$119~$\mu$m.
The $\Lambda$--doubling lines are  only resolved in the FP spectrum 
(Fig.~\ref{varias_molec}d).
Stacey et al. (1987) detected the CH $\sim$149~$\mu$m lines
with KAO only towards Sgr~B2(M). ISO observations show that CH
is present in the whole region.
Furthermore, the line profile obtained with the FP towards Sgr~B2(M)
is very similar to that 
observed with KAO (at a resolution of 62~km~s$^{-1}$ and a beam of 55$''$).
The lines have the same broad profiles seen in fundamental lines of
H$_2$O and OH  produced by the
foreground clouds not associated with Sgr~B2.
Two distinct absorption peaks at $\simeq$0 and $\simeq$+50~km~s$^{-1}$
are clearly detected. 

For each position in the cross--like map, the CH column density in the
ground--state has been estimated considering a single unresolved 
rotational line. The rotational line strength
we have used is the sum of each  $\Lambda$--doubling line strength
calculated from the individual hyperfine transitions. 
The absorption across the extended region is proportional to 
the total CH column density. We found column densities for 
the $^2\Pi_{1/2}$ J=1/2 ground level
between  $(0.8-1.8)\cdot10^{15}$~cm$^{-2}$ with the larger
values centered around Sgr~B2(M,N). 
Using the FP observations (Fig.~\ref{varias_molec}d),
we derive a column density of $1.7\cdot10^{15}$~cm$^{-2}$ for
Sgr~B2(M), similar to the 
value obtained by Stacey et al (1987). 
We have also searched for several CH excited rotational lines  
at $\sim$180~$\mu$m, but opposite to OH no other lines have detected
(see Cernicharo et al. 1999  for undetected FP lines spectra). 
The lack of absorption from other lines of CH rather than those 
connecting with the ground level suggest that the molecule is only 
abundant in the foreground  clouds and in the external layers 
of Sgr~B2  where collisional excitation is unimportant. Thus, the above column
densities can be a good approximation to the total CH column density.
Only higher spectral resolution maps of the CH $\sim$149 $\mu$m line 
will allow the accurate division of these column densities
into the different foreground clouds. In the case of 
the Sgr~B2(M) line of sight, only $\sim30\%$ of the total
column density  arises from Sgr~B2
(Stacey et al. 1987).
The lack of CH detections in excited rotational states confirms 
that this species predominantly appears in low--density clouds.
Due to the large continuum opacity at $\sim$149~$\mu$m
and $\sim$180~$\mu$m, the CH emission 
coming from the inner and denser regions does not
contribute much to the far--IR observations. Such CH emission
is observed at radio wavelengths (Stacey et al. 1987).

\subsection{Far--IR detection of NH$_2$ and NH}

In this section we report the far--IR detection of amidogen, NH$_2$, and
imidogen, NH, key radicals toward Sgr~B2(M). It is the first time that the
ortho--NH$_2$ species have been observed in the ISM. We also analyze
in more detail the first detection of NH in the dense ISM
(Cernicharo, Goicoechea \& Caux 2000). 

NH$_2$ is a light and floppy molecule thought to be an important reaction 
intermediate in the production/destruction of N--bearing molecules such as 
ammonia. NH$_2$ is an asymmetrical molecule with a $^2$B$_1$ ground 
electronic state characterized by an intricate rotational spectrum.
Because of the 2 equivalent H nuclei, ortho and para modifications
can be distinguish. The spin--rotation interaction caused by the unpaired 
electron splits each rotational level into two sublevels which are further
split by hyperfine interactions caused by the $^{14}$N nuclear spin.
Additional splitting occur in the ortho-- levels due to the resultant
proton spin.
Although its presence as a photodissociation product of NH$_3$
in cometary spectra have been known since early 1940's 
(Swings, McKellar \& Minkowski  1943) its interstellar detection
had to wait 50 years until the detection of the 
1$_{10}$--1$_{01}$ lines of para--NH$_2$ at millimeter
wavelengths towards Sgr~B2 (van Dishoeck et al. 1993).
This is the only detection of the molecule reported in the ISM.
Several far--IR spectral features have been recently
observed in the laboratory (Gendriesch et al. 2001 and references
therein). Here we report the ISM detection of some of them.

\begin{figure}
\centering
\includegraphics[angle=0, width=12cm]{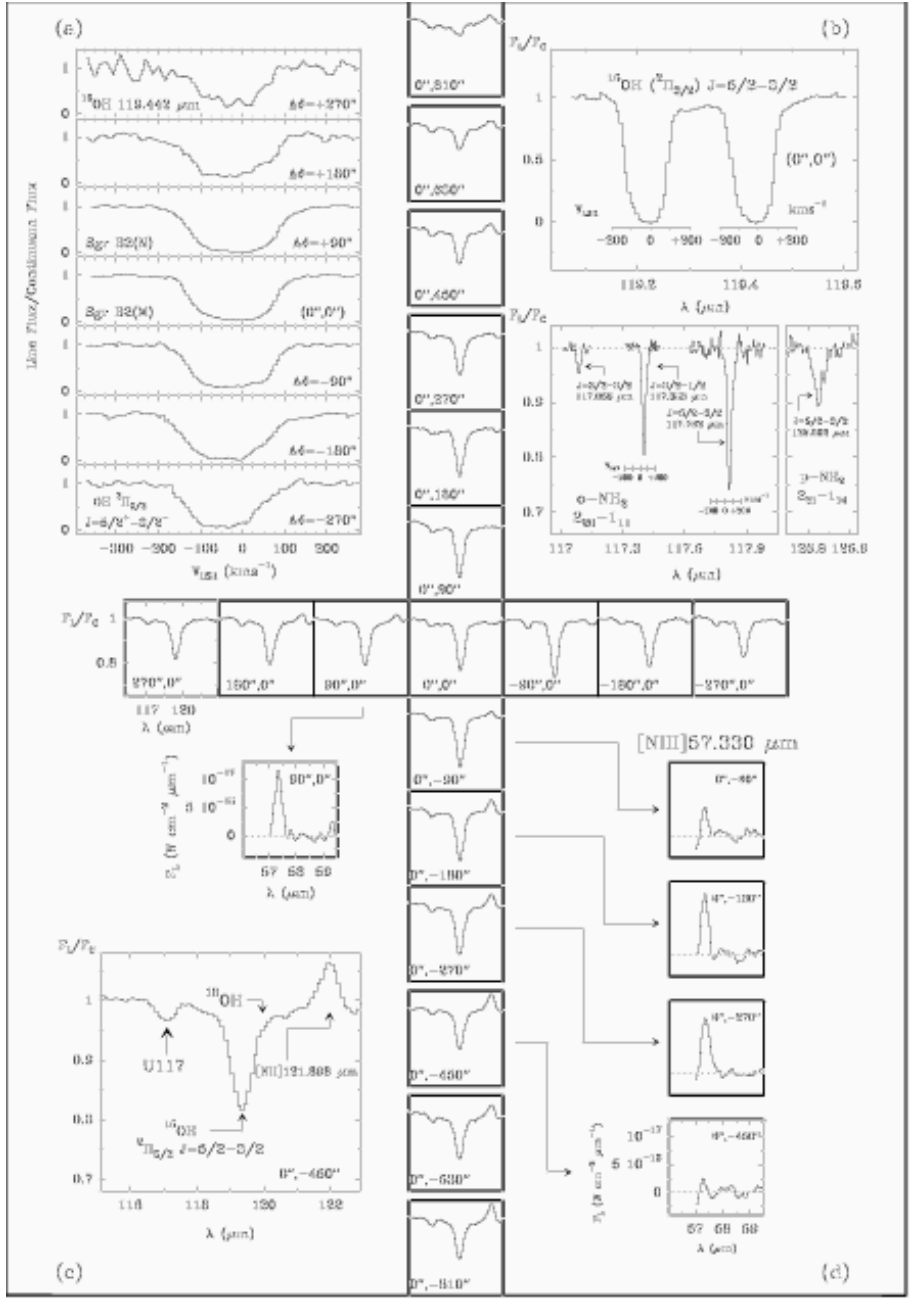}
\figcaption{\small Raster map obtained with the LWS grating between
115--121 $\mu$m. In each box, offset positions are given in arcsec
respect to the $(0'',0'')$ central position at:
$\alpha= 17^h44^m10.61^s$, $\delta=-28^o22^{'}30.0^{''}$ [J1950]. The 
intensity scale corresponds to the continuum normalized flux and the
abscissa to the wavelength.
(a) LWS-FP declination raster of the $^{16}$OH 119.442~$\mu$m line.
(b) $Top$: Fundamental OH $\Lambda$--doublet detected with the LWS-FP in
$(0'',0'')$ around $\sim$119~$\mu$m.
$Bottom$: ortho and para NH$_2$ lines detected with the LWS-FP in $(0'',0'')$
around $\sim$117 and 126~$\mu$m. The NH$_2$  $\sim$117 triplet may contribute
to the U117 line observed at much lower resolution.
(c) Main features of the grating raster map labelled. 
(d) Raster map positions with clear [\NIII]57~$\mu$m detections.
\label{Figuron-OH}}
\end{figure}

Figure \ref{Figuron-OH}b (lower box) shows the three spin--rotational 
components of o--NH$_2$ 2$_{20}$--1$_{11}$ transition
at $\sim$117~$\mu$m, while 
only one spin--rotational 
component of the para--NH$_2$ 2$_{21}$--1$_{10}$ transition
at $\sim$126~$\mu$m has been detected. No more low excitation NH$_2$ 
rotational lines  with E$_l$$<$100 K have been found. 
The lines are centered at Sgr~B2 velocities, without absorption at 
negative velocities from the foreground clouds.
The different spin--rotational line strengths 
were calculated by adding the individual hyperfine structure line
strengths that are listed in \textit {Cologne Database for Molecular 
Spectroscopy} (CDSM) catalog for NH$_2$ (M\"{u}ller et al. 2001).
However, ortho-- and para-- species were analyzed as two different
molecules and  the energy  of the para--levels were referred to the
lowest para--level that we assign to 0 K energy.
The two different spin--rotational partition functions
were then computed. 
The theoretical line strength ratio for the three 
ortho--NH$_2$ 2$_{20}$--1$_{11}$  spin-rotational lines 
is $10.0/5.6/1.1$  (S$_{117.8}$$/$S$_{117.4}$$/$S$_{117.1}$)
while the optical 
depth ratio derived from FP observations is $10.0/7.3/1.7$.
This suggests that the $J=5/2-3/2$ line (117.792~$\mu$m)
is moderately thick. With this ratio we can compute the
total opacity of the rotational transition and estimate the ortho--NH$_2$
column density. Assuming a Boltzmann population of the rotational
levels and excitation temperatures between 20 and 30 K
(T$_{ex}$$\leq$T$_d$) we found $N$(o-N$H_2$)$=(1.2\pm0.3)\cdot10^{15}$~cm$^{-2}$. 
We have only detected the more intense spin--rotational
component of the para--NH$_2$ 2$_{21}$--1$_{10}$ transition
at $\sim$126.8~$\mu$m. The theoretical line strength ratio for the three lines 
is $10.0/5.5/1.1$ (S$_{126.8}$$/$S$_{126.4}$$/$S$_{126.0}$).
Due to the high continuum level measured by the LWS/FP in the TDT 50601112
at $\sim$126~$\mu$m, and the fact that only three scans were available
(low S/N ratio), the other two components are expected to be under the 
detection limit of the AOT L03 observations. No definitive assignation
have been done.
Assuming the the component at $\sim$126.8~$\mu$m is optically thin,
we derive $N$(p--NH$_2$)$\sim4 \cdot10^{14}$~cm$^{-2}$.
Van Dishoeck et al., derived $N$(p--NH$_2$)=$
(1.3\pm0.3)\cdot10^{15}$~cm$^{-2}$ from escape probability calculations.
Hence, this value has to be considered as 
an upper limit to the para-- column density if both the 
submm (ground--state) and far--IR (excited) lines  
arises from the same gas. 
It is also possible that far--IR NH$_2$ rotational
lines trace a warmer component of lower column density of material.
Taking into account both observations,
we estimate a total (ortho-- + para--) NH$_2$ column density of
(1.5--3.0)$\cdot$10$^{15}$~cm$^{-2}$.
The upper limit reflects the para--NH$_2$ column density derived
from submm observations, while the lower limit is  an
estimation of the $N$(p--NH$_2$) based on far--IR observations.

 We have finally explored the possibility that the unidentified
feature U117 ($\sim$117.1~$\mu$m) observed in all positions of the 
raster map (see Fig.~\ref{Figuron-OH}) arises from the three 
spin--rotational components of the ortho--NH$_2$ 2$_{20}$--1$_{11}$ 
transition observed at higher resolution towards Sgr~B2 (M) 
(Fig.~\ref{Figuron-OH}b).
In that case, the widespread absorption will correspond
to a nearly constant  ortho--NH$_2$ column density of 
$\sim$2$\cdot$10$^{15}$~cm$^{-2}$. However, the unresolved NH$_2$ line
in the gratings should be centered at $\sim$117.6~$\mu$m, which
prevents from a definitive assignation to NH$_2$. In addition,
we have studied  other possible molecular contributors with 
significant line strength.  In particular, the U117 line
could also arise from several 
$b-type$ transitions of slightly
asymmetrical species such as  HNO, HNCO, or HOCO$^+$, or it can be
the $Q$--branch of a low energy bending mode of a carbon chain.
As a example, theoretical calculations predict a 117--125~$\mu$m wavelength  
for the  $\nu_{11}$ bending mode of  C$_7$, 
but no laboratory bands have been still attributed 
(Kurtz \& Adamowicz 1991; Martin et al. 1995).

Also related with the formation/destruction of the widely observed
NH$_3$ molecule is the NH radical. 
The NH (X$^3\Sigma^{-}$) rotational spectrum is also complicated by
the different angular couplings between the rotation, the electronic spin,
and  the H and $^{14}$N nuclear spin momenta 
(cf. Klaus, Takano \& Winnewisser 1997). 
Before our far-IR detection in Sgr~B2,
NH detections had been only reported in the diffuse and translucent clouds 
(Meyer \& Roth 1991). The enhanced NH column 
densities found in these environments have been used to support 
that  N--chemistry is dominated by grain--surface reactions
instead of gas-phase reactions (Wagenblast et al. 1993). \\\\

The LWS/FP can only resolve different spin--rotational components
of a rotational line.
The two  detected lines (Fig.~\ref{varias_molec}a) arise from 
the $N=2\leftarrow1$ rotational transition with the lower energy level
at $\sim$45~K. These lines peak at Sgr~B2 velocities.
The $N=2\leftarrow1$ lines arise from Sgr~B2(M) and represent
the first detection of this species in a dense molecular cloud.
Future observations of the  $N=1\leftarrow0$ transition from the
fundamental ground--state at
$\sim$1 THz will provide the signature of the NH present in the diffuse
molecular clouds of the line of sight and will complement the observations
of NH in translucent clouds through its electronic spectrum.
 We have computed the spin--rotational line strengths from the 
individual hyperfine structure line strengths listed in the 
\textit{JPL catalog} for NH (Pickett et al. 1998).
A similar analysis of fine structure to that 
done with NH$_2$ was carried out to extract the opacity from the NH 
2$_3$-1$_2$/2$_2$-1$_1$ optical depth ratio. 
Assuming an excitation temperature of 20--30~K we derive 
$N$(NH)=(4$\pm$2)$\cdot$10$^{14}$~cm$^{-2}$.

\subsubsection{Nitrogen chemistry and shocks}

The ISO observations of Sgr~B2(M) have given the opportunity of
observing simultaneously the NH$_3$, NH$_2$ and NH species. 
These molecules represent the best signature of the prevailing N--chemistry
in the outer regions of the cloud. Far--IR NH$_3$ lines have been 
analyzed  by Ceccarelli et al. (2002). Their LVG
calculations showed the NH$_3$ column density in the warm absorbing layers 
is $(3.0\pm1.0)\cdot10^{16}$~cm$^{-2}$. 
Assuming that NH$_3$ far--IR and radio (H\"{u}ttemeister et al. 1995)
lines arise in the same region, 
they derived large  temperatures (T$_k\sim$700~K), similar to those 
obtained from radio observations.
However, the ammonia observed in absorption
against the radio continuum is dominated by 
the  molecular gas in front of the HII regions. Due to the
large opacity in the far--IR, the bulk of the gas sampled by ISO  
only refers to the Sgr~B2 envelope embedding the star formation
regions.

There is strong evidence that both the NH$_3$ heating  and 
the observed metastable column densities can be reproduced if grain--surface
formation and sputtering by low--velocity shocks are taken into account
(Flower et al. 1995). In fact, the same shock models satisfactory
explain the recent NH$_3$ (Ceccarelli et al.)
and NH$_2$ (this work) column densities derived from
ISO observations.

The special  formation conditions of gas phase NH$_3$
(grain chemistry and  mantle erosion) and its survival conditions 
(easily photodissociated) suggests that studies of the 
neutral gas in Sgr~B2  by means of the NH$_3$ absorption
are only sensitive to its specific conditions.\\\\

We found that the observational NH$_3$/NH$_2$/NH$\simeq$100/10/1 column 
density ratio can not be explained in terms of dark cloud models, these
predict  NH$_3$/NH$_2$$<$3  (Millar et al. 1991) and this ratio
would be even smaller if photodissociation of NH$_3$ is included.
In addition, Sternberg \& Dalgarno (1995) derived a  
NH$_3$/NH column density ratio  $>$2$\cdot10^3$ in the region where
the incident far--UV field is completely attenuated.
A  PDR contribution to the observed N--bearing radicals may be possible
as photodissociation products of NH$_3$. 
However, current models predict 
NH$_2$/NH$<$1 and NH$_3$/NH$<$1 column density ratios in the regions
affected by the far--UV radiation field (A$_V$$<$5~mag; Sternberg \& 
Dalgarno 1995), which are not observed at least in the averaged picture 
given by  the large ISO/LWS beam.
The large column densities of warm ammonia
found in the Sgr~B2 envelope  (Ceccarelli et al. 2002) can still
be compatible with its photodissociation if a notable
enhancement in the NH$_3$ grain--surface formation and an
efficient mantle erosion mechanism dominates the ammonia chemistry.
Such mechanism are the widespread low--velocity shocks that
liberate large amounts of NH$_3$ from the
grains and also heat the NH$_3$ in the gas phase.

The chemistry in Sgr~B2 is a challenging issue as
the same shock models that apparently
reproduce the N--chemistry (Flower et al. 1995)
fail in reproducing the O--chemistry traced by far--IR observations.
The predicted H$_2$O column density is almost two orders of magnitude
larger than the observed by ISO 
(Goicoechea \& Cernicharo 2001; Cernicharo et al. 2004, in prep.)
while the predicted OH column density is an
order of magnitude lower than the observed (GC02). 
Thus, the large OH abundance
seems more related with the photodissociation of H$_2$O.
Recent analysis of the extended H$_2$O $1_{10}-1_{01}$ absorption
observed by SWAS around Sgr~B2 (Neufeld et al. 2003) and of the HDO absorption
toward Sgr~B2(M,N) (Comito et al. 2003) support the above (ISO) column
densities for the water vapor.
Finally, the OH absorption is correlated
with the warm [\OI] emission and this could be related with the OH
photodissociation in the outer layers.

\section{Summary}

The present far--IR continuum and spectral observations of the Sgr~B2
region have revealed a new perspective of the less known extended envelope
of the complex. The ISO observations show the presence of an
extended component of ionized gas reaching very large 
distances from the regions of known massive star formation.
Photoionization models show that  high effective 
temperatures are possible if the ionization parameter is low.
We found that the radiation can be  characterized by a hard ionizing continuum typical of
an O7 star (T$_{eff}$$\simeq$36000~K) and the ionization of the Sgr~B 
complex is dominated by Sgr~B2.\\

We suggest that the whole region must be highly clumped and/or
fragmented, so that the diffusion of the far--UV radiation field
allows the  ionized /neutral (warm) /neutral (cold and dense) material
to exist throughout the cloud.
However, the exact three dimensional location of the
ionization sources relative to the extended cloud is not clear, 
but it also determines the regions preferably illuminated by the UV 
radiation (southern and
eastern regions). It is plausible that the moderate density cloud
around Sgr~B2(M,N) blocks the ionization radiation
in the northern and western directions. Another possibility
is the presence of evolved stars and/or young massive stars in the 
envelope of Sgr~B2 itself, so that the  UV field on the extended
cloud is the averaged interstellar field.
The observed X-ray emission could also play a
role in the large scale ionization, but the effects on the neutral gas
are more difficult to determine.
In any case, we have presented observational evidence that
gas is photochemically active far from the ionizing sources.
This is reflected in the neutral gas heating  and in the column
densities of some molecules.

In addition, molecular tracers and atomic fine structure
tracers do not show evidence
of high--velocity shocks. Hence, the ionized gas can not be explained
in terms of high--velocity dissociative shocks. 
It seems that the well established
widespread low--velocity shocks (H\"{u}ttemeister et al. 1995;
Mart\'{\i}n--Pintado et al. 1997)
are not the only mechanism heating the gas to
temperatures larger than those of the dust.
Coexistence of mechanical and radiative--type  heating mechanisms based on the
effects of a UV radiation field that permeates an inhomogeneous 
medium seems to be the rule in Sgr~B2  and also in the bulk of 
GC molecular clouds observed by ISO 
(Rodr\'{\i}guez--Fern\'andez et al. 2004, in prep.).

According to the extended distribution of molecular species
such as H$_2$O, OH and CH, and the large column density
of key molecular species detected in the Sgr~B2(M) position,
Sgr~B2 is one of the richest and peculiar clouds in the galaxy.

The geometrical properties
of Sgr~B2 (clumped structure, extended envelope, centrally 
condensed hot cores and compact HII regions); its physical conditions
(high average densities, widespread warm gas, cool dust, turbulence
and enhanced interstellar radiation field);
and its chemical complexity 
(some molecules detected only toward Sgr~B2 and nowhere else in the Galaxy,
extended emission  of refractory molecules, etc.) mimics 
a miniature Galactic Center with only $\sim$15$'$ extend.
Therefore, Sgr~B2 provides a good $\it{template}$ to study the mean
physical and chemical processes in a galactic nucleus with enough spatial
resolution, from the developing clusters of hot stars,
to the regions without significant luminous internal heating sources,
but exposed to the mean GC environment conditions.

The actual far--IR spectral and spatial resolution makes a similar analysis
of other  galactic nuclei rather speculative. However, it seems 
tempting that extrapolating the far--IR spectrum of $\approx$10$^5$ Sgr~B2 
clouds to a distance of a few Mpc will yield a similar  spectrum to 
those  existing of normal and IR galaxies
such as Arp 220.
Recent application of PDR models to different samples of galaxies observed
in the far--IR show that the physical conditions of extragalactic PDRs
do not differ much from the values derived for Sgr~B2
(WHT90; Malhotra et al. 2001). Hence, Sgr~B2 also plays a template
role for extragalactic ISM studies.

To summarize, we have carried out a  medium resolution ($\sim$1000~km~s$^{-1}$)
mapping from 43 to 197~$\mu$m of the Sgr~B2 region (9$'$~$\times$~27$'$)
and have observed the Sgr~B2(M) central source at high
resolution ($\sim$35~km~s$^{-1}$). The main conclusions of this 
work are:

\begin{itemize}

\item The far--IR spectra show an extended region of strong dust emission
(L$_{FIR}\simeq10^7L_{\odot}$). The observed continuum emission
is best fitted with a dust component with a temperature
of 13--22~K and a warmer component with a temperature of 24--38~K.
The warmer component contributes in less than 10\% to the total optical
depth.

\item The [\OIII], [\NIII] and  [\NII] fine structure line emission 
has revealed an extended component of ionized gas. The averaged 
electronic density is $\sim$240~cm$^{-3}$.
The ionizing radiation can be characterized by T$_{eff}\simeq36000$~K 
but a low ionization parameter. The total number of 
Lyman photons  needed to explain such component is approximately equal to
that of the  HII regions within Sgr~B2. The southern regions of the Sgr~B
complex may  be also influenced by additional ionizing sources near 
the location of Sgr~B1.

\item The location and geometry distribution of the ionizing 
sources relative to the extended cloud favors the ionization
of the southern and eastern regions of Sgr~B2.
Also the  clumpy structure of gas in Sgr~B2 that surrounds
the ionizing sources 
allows the radiation to penetrate large distances throughout the envelope,
so that PDRs  are numerous  throughout the cloud
at the interface of 
the ionized and the neutral gas.

\item Comparison of the [\OI] and [\CII] lines 
with PDR models indicates a G$_0$$\simeq$10$^{3-4}$ enhancement of the far--UV 
field and a n$_H$ density of 10$^{3-4}$~cm$^{-3}$ in these PDRs. 

\item Extended photoionization and photodissociation are also taking
place in Sgr~B2 envelope in addition to other processes
such as widespread low--velocity shocks. The origin of the rich chemistry 
is  probably a result of the combination of both scenarios.

\item The ground--state rotational transitions of light hydrides such
as OH, CH and H$_2$O produce absorption in all observed
positions representing evidences that these far-IR lines 
are widespread in molecular clouds.

\item Radiative transfer models of OH show that the neutral gas
in the external molecular layers have physical conditions 
($n_{H_2}$=10$^{3-4}$~cm$^{-3}$; T$_k$$\gtrsim$40--100~K)
in between those derived from the ionized gas and those of the denser
gas in the star forming cores.

\item 
The LWS/FP spectrum of Sgr~B2(M)
shows the most abundant species that could be detected with ISO
in the far--IR. The H$_2$O/OH/O$^0$ chemistry in the envelope seems 
dominated by photodissociation processes while 
the NH$_3$/NH$_2$/NH$\simeq$100/10/1  column density ratios are still better
explained under low--velocity shocks activity.\\

\end{itemize}

Future far--IR heterodyne instruments, such as HIFI on board the 
$\it{Herschel}$  $\it{Space}$  $\it{Observatory}$ will observe the
Sgr~B2 region and other galactic nuclei with larger spatial and spectral
resolution than ISO.
A direct observation of the inhomogeneous nature of the ionized gas,
the warm PDRs and the shocked regions will then be possible.

\acknowledgments
      We thank Spanish DGES and PNIE for funding
support under grant ESP98--1351--E, PANAYA2000-1785,
AYA2000-1888-E, AYA2000-1974-E, and ESP2001-4516
We thank J.~Mart\'{\i}n--Pintado for stimulating discussions about 
Sgr~B2, M.A.~Gordon for providing us the IRAS maps of Sgr~B2 and 
J.~Fisher and E.~G\'onzalez-Alfonso for useful suggestions.
We also thank M.A.~Orlandi for help in the data analysis and
the anonymous referee for his/her useful comments
which resulted in an improvement of the paper.
NJR-F has been supported by a Marie Curie Fellowship of the European
Community program ``Improving Human Research Potential and the
Socio-economic Knowledge base'' under contract number HPMF-CT-2002-01677.

\clearpage
         
\clearpage

\clearpage

\clearpage

\clearpage

\clearpage

\clearpage

\clearpage

\clearpage


\begin{deluxetable}{crrrrrrr}
\tabletypesize{\scriptsize}
\tablecaption{Fluxes derived from gaussian fits to the 
fine structure lines observed with the LWS grating. 
Fluxes are in units of 10$^{-19}$~W~cm$^{-2}$
and numbers in parentheses are the statistical errors.
Offsets (in arcsec) respect to the map center, Sgr~B2(M). \label{gr-fluxes}}
\tablewidth{0pt}
\tablehead{ 
  & \colhead{[\CII]} & \colhead{[\OI]} & \colhead{[\NII]} & \colhead{[\OIII]} & 
\colhead{[\OI]}  &  \colhead{[\NIII]}  & \colhead{[\OIII]} \\

\colhead{$\lambda$($\mu$m)} & \colhead{157.741}  & \colhead{145.525} &
\colhead{121.898} & \colhead{88.356}  & \colhead{63.184} 
&  \colhead{57.317}  & \colhead{51.815}  \\
\colhead{map position} & & & & & & &   
}
\startdata
(0,810) & 68.0(2.3)  & 5.1(0.1)  & 5.6(0.5)  & 2.8(1.2)   & 20.1(1.2)  & $<$4        & $<$7       \\
(0,630) & 69.4(2.5)  & 4.1(0.4)  & 4.8(1.1)  & 4.0(1.5)   & 21.3(1)    & $<$3        & 7.8(4.1)   \\
(0,450) & 86.0(4.1)  & 6.4(0.6)  & 10.9(1.7) & 5.9(0.2)   & 31.2(1.5)  & $<$1        & 8.6(2.2)   \\
(0,270) & 79.0(5.8)  & 8.4(3.3)  & 3.0(1.0)  & 4.1(0.9)   & 33.0(1.2)  & $<$4        & $<$5        \\
(0,180) & 62.6(5.9)  & 10.3(6.0) & 3.4(0.5)  & 4.3(1.0)   & 20.6(1.5)  & $<$4        & $<$6         \\
(0,90)  & ---$^a$    & 42.0(1.3) & $<$35     & 28.4(4.2)  &-14.9(2.9)  & $<$11       & 8.2(3.1)    \\
(0,0)   & ---$^a$    & 73.3(8.2) & $<$44     & 41.5(17.2) &-137.2(18.1)& $<$17       & 21.9(6.2)   \\ 
(0,-90) & 149.0(10.5)& 39.9(1.0) & 18.5(4.6) & 71.5(8.3)  & 25.6(2.6)  & 8.1(2.4)    & 26.3(4.7)   \\ 
(0,-180)& 237.2(8.8) & 29.7(3.3) & 41.6(6.0) & 98.0(4.7)  & 83.7(3.6)  & 21.7(2.9)   & 60.0(7.1)   \\ 
(0,-270)& 235.3(6.7) & 24.7(2.1) & 33.9(3.6) & 67.4(6.7)  & 73.2(2.7)  & 25.5(2.9) & 25.9(3.4)  \\  
(0,-450)& 210.7(4.0) & 14.9(1.4) & 39.6(4.6) & 19.9(3.2)  & 63.6(1.4)  & 11.1(3.5) & 22.8(7.9)  \\
(0,-630)& 146.6(4.5) & 11.1(0.5) & 17.6(3.6) & 10.5(1.5)  & 31.0(3.8)  & 5.3(3.6)  & 20.6(10.6)  \\
(0,-810)& 113.8(2.0) & 6.8(0.5)  & 16.6(1.0) & 15.0(1.4)  & 23.3(1.7)  & 3.3(1.2)  & 9.2(1.1)    \\
(270,0) & 104.3(6.0) & 13.6(1.7) & 9.1(1.6)  & 6.3(1.5)   & 38.9(2.6)  & $<$4      & $<$1        \\
(180,0) & 144.7(5.6) & 11.1(1.1) & 19.0(2.0) & 18.1(1.6)  & 46.9(2.6)  & $<$7      & 18.1(2.9)   \\
(90,0)  & 174.2(7.8) & 32.5(2.1) & 35.0(2.8) & 67.9(4.5)  & 45.9(2.0)  & 28.1(3.0)  & 26.1(4.0)   \\ 
(-90,0) & 60.6(8.7)  & 21.4(2.4) & $<$8      & 13.4(3.4)  & 12.6(1.9)  & $<$3        & 5.1(8.5)   \\
(-180,0)& 108.0(4.9) & 10.0(1.1) & $<$7      & 11.0(7.0)  & 23.3(1.8)  & $<$3        & 3.9(1.9)  \\               
(-270,0)& 94.8(5.0)  & 9.8(0.6)  & $<$6      & 9.7(1.3)   & 17.8(1.3)  & $<$4        & 3.4(0.3)   \\

\enddata

\tablenotetext{a}{At the low resolution of the LWS/grating, the
[\CII]158 $\mu$m line is not detected in Sgr~B2(M,N). When observed
at higher resolution (see Fig.~\ref{ionic}b) the resolved line profile shows an
absorption feature produced by the foreground gas that reduce the 
expected line flux in the grating to the noise level of the spectra.
Hence, the non detection in the grating is consistent with the
detection in the FP.}

\end{deluxetable}

\clearpage

\begin{deluxetable}{cccc}
\tabletypesize{\scriptsize}
\tablecaption{Dust temperatures derived by fitting the far--IR
continuum emission with 2 gray bodies and luminosities in the LWS range.
\label{tab-Tdust}}
\tablewidth{0pt}
\tablehead{ 
\colhead{position$^a$} & \colhead{T$^{warm}_{dust}$(K)} &
\colhead{T$^{cold}_{dust}$(K)}
& \colhead{L$_{LWS}$(L$_{\odot}$})
}
\startdata
(0,810)   & 30-33   &   17-20 &   6.91E4  \\
(0,630)   & 30-33   &   17-20 &   9.57E4  \\      
(0,450)   & 31-35   &   18-22 &   1.37E5  \\
(0,270)   & 25-29   &   13-20 &   1.86E5  \\
(0,180)   & 26-35   &   16-20 &   2.56E5  \\ 
(0,-90)   & 27-30   &   17-19 &   6.20E5  \\  
(0,-180)  & 32-36   &   18-21 &   4.74E5  \\ 
(0,-270)  & 34-38   &   18-22 &   3.61E5  \\
(0,-450)  & 36-38   &   17-20 &   3.17E5  \\
(0,-630)  & 32-34   &   18-21 &   1.91E5  \\ 
(0,-810)  & 34-37   &   18-22 &   1.20E5 \\
(270,0)   & 25-27   &   14-19 &   2.63E5  \\
(180,0)   & 29-32   &   17-20 &   2.90E5  \\
(90,0)    & 31-34   &   18-20 &   5.21E5  \\
(-90,0)   & 24-26   &   15-19 &   5.07E5  \\ 
(-180,0)  & 24-26   &   13-19 &   2.60E5  \\
(-270,0)  & 24-25   &   13-19 &   2.24E5  \\
\enddata

\tablenotetext{a}{Offsets (in arcsec) respect to Sgr~B2(M).}

\end{deluxetable}

\begin{deluxetable}{crrrrrr}
\tabletypesize{\scriptsize}
\tablecaption{Selected line ratios after correcting for the two
limits of visual extinction (A$_V^{min}$ and A$_V^{max}$) 
derived for best--fit gray body models ($\beta$=1.0--1.5). 
The different beam sizes of each LWS detector are taken 
into account and  extended emission is considered. \label{tab-ratios}}
\tablewidth{0pt}
\tablehead{ 
 & \colhead{warm dust} & \colhead{cold dust} & \colhead{[\OIII]} & 
 \colhead{$n_e$([\OIII])} & \colhead{[\NIII]/[\NII]} &
 \colhead{T$_{eff}$} \\
 
\colhead{map position$^a$} & \colhead{A$_V$ (mag)} &
\colhead{A$_V$(mag)} 
& \colhead{R(52/88)} & \colhead{log(cm$^{-3}$)} & \colhead{R(57/122)} &
\colhead{($\div$10$^3$ K)}
}
\startdata
(0,810) & 1.2-1.5  & 15-55   & $<$2.40     & $<$3.18    & $<$0.72     & $<$33.2 \\
(0,630) & 1.6-2.0  & 23-84   &  1.79-2.23  &  2.83-3.10 & $<$0.72     & $<$33.2 \\
(0,450) & 1.1-1.8  & 25-92   &  1.36-1.73  &  2.49-2.79 & $<$0.26     & $<$31.8 \\
(0,270) & 5.2-8.0  & 41-112  & $<$1.61     & $<$2.70    & $<$1.77     & $<$35.6 \\
(0,180) & 0.7-1.7  & 131-294 & $<$3.55     & $<$3.66    & $<$2.73     & $<$35.2 \\
(0,-90) & 16-18    & 367-877 &  1.27-8.11  &  2.41-4.67 &  1.60-11.3  & (35.0-37.2) \\
(0,-180)& 3.7-5.2  & 148-493 &  0.90-3.18  &  1.99-3.53 &  0.78-2.95  & (34.9-35.3) \\
(0,-270)& 2.8-3.8  & 59-205  &  0.41-0.69  &  1.03-1.67 &  0.79-1.39  & (35.0-35.8) \\
(0,-450)& 3.2-5.3  & 91-249  &  1.47-2.64  &  2.59-3.30 &  0.35-0.64  & (32.2-33.1) \\
(0,-630)& 3.5-3.9  & 23-85   &  1.83-2.30  &  2.85-3.13 &  0.28-0.36  & (31.9-32.2) \\
(0,-810)& 1.1-1.3  & 16-59   &  0.55-0.64  &  1.39-1.58 &  0.18-0.21  & (32.8-33.1)  \\ 
(270,0) & 27-28    & 156-536 & $<$1.19     & $<$2.24    & $<$3.13     & $<$36.6 \\ 
(180,0) & 4.9-6.6  & 78-276  &  1.14-2.38  &  2.28-3.18 & $<$0.95     & $<$33.6 \\ 
(90,0)  & 7.4-9.0  & 168-565 &  0.62-2.64  &  1.53-3.30 &  1.31-6.07  & (35.7-36.3) \\
(-90,0) & 28-34    & 228-579 &  0.85-2.97  &  1.92-3.44 &  ---        & --- \\
(-180,0)& 21-26    & 62-168  &  0.43-0.63  &  1.10-1.55 &  ---        & ---\\
(-270,0) & 21-25   & 45-124  &  0.38-0.50  &  0.94-1.27 &  ---        &  ---\\

\enddata

\tablenotetext{a}{Offsets (in arcsec) respect to the map center,
Sgr~B2(M).}

\end{deluxetable}

\clearpage

\begin{deluxetable}{ccrr}
\tabletypesize{\scriptsize}
\tablecaption{Published Far--IR features detected with the LWS/FP
toward Sgr~B2(M) between 181.1-119.6~$\mu$m
\label{survey1}}
\tablewidth{0pt}
\tablehead{ 
\colhead{species} & \colhead{transition} & \colhead{$\lambda_{rest}$($\mu$m)} 
& \colhead{references}
}
\startdata
H$_3$O$^+$     & $1_1^- - 1_1^+$      &  181.05 & (5) \\
H$_{2}^{18}$O  & $2_{12}-1_{01}$    & 181.05 & (2),(5),(11) \\
H$_2$O       & $2_{21}-2_{12}$    & 180.49  & (11)   \\
H$_2$O       & $2_{12}-1_{01}$    & 179.53  &  (2),(11)   \\
H$_2$O       & $3_{03}-2_{12}$    & 174.63 &    (2),(11)  \\
NH$_3$       & $(3,2)a-(2,2)s$    &  169.97 &   (7)  \\
C$_3$        & $P(4)\hspace{0.2cm}\nu_2=1-0$ & 167.68  &   (4)  \\
NH$_3$       & $(3,2)s-(2,2)a$  & 165.60  &   (7)  \\
OH         & $^2\Pi_{1/2}\hspace{0.2cm} J=3/2^--1/2^+$ & 163.40  &   (8)  \\
OH         & $^2\Pi_{1/2}\hspace{0.2cm} J=3/2^+-1/2^-$ & 163.12  &   (8)  \\
C$_3$        & $Q(4) \hspace{0.2cm}\nu_2=1-0$ & 158.06  &  (4)  \\
C$^+$        & $^2P_{3/2}-{^2P}_{1/2}$  & 157.74  &  (1),(10)  \\
C$_3$        & $Q(6)\hspace{0.2cm} \nu_2=1-0$   & 157.26  &  (4)  \\       
H$_2$O       & $3_{22}-3_{13}$    & 156.19  &  (11)   \\      
C$_3$        & $Q(8)\hspace{0.2cm} \nu_2=1-0$   & 156.19  &  (4) \\
C$_3$        & $Q(10)\hspace{0.2cm} \nu_2=1-0$  & 154.86  &   (4)  \\       
NH           & $2_3-1_2$          & 153.34  &   (4)  \\
C$_3$        & $Q(12)\hspace{0.2cm} \nu_2=1-0$  & 153.30  &   (4) \\
NH           & $2_2-1_1$          &  153.10 &   (1)  \\
C$_3$        & $R(2)\hspace{0.2cm} \nu_2=1-0$   & 152.29  &   (4),(16)  \\   
CH           & $^2\Pi_{3/2}-{^2\Pi}_{1/2} \hspace{0.2cm} J=3/2^--1/2^+$ &149.39 &(1),(13)  \\
CH           & $^2\Pi_{3/2}-{^2\Pi}_{1/2} \hspace{0.2cm} J=3/2^+-1/2^-$ &149.09 &(1),(13)   \\ 
C$_3$        & $R(4)\hspace{0.2cm} \nu_2=1-0$ &148.04 &  (4)  \\ 
O$^0$        & $^3P_0-{^3P}_1$ & 145.53 &  (1),(10)\\ 
C$_3$        & $R(6)\hspace{0.2cm}  \nu_2=1-0$ &143.88 &  (4)\\ 
H$_2$O       & $3_{13}-2_{02}$ &138.53 &  (11)\\ 
NH$_3$       & $(4,3)a-(3,3)s$   & 127.11 &  (7)\\ 
NH$_2$       & $2_{21}-1_{10} \hspace{0.2cm} J=5/2-3/2$ & 126.80 &  (1)\\ 
NH$_3$       & $(4,2)s-(3,2)a$ & 124.80 &  (7)\\ 
NH$_3$       & $(4,3)s-(3,3)a$ & 124.65 &  (7)  \\ 
HF           & $J=2-1$ & 121.70&  (3)\\
$^{18}$OH    & $^2\Pi_{3/2} \hspace{0.2cm} J=5/2^+-3/2^-$ & 120.17 &(8),(14)\\ 
$^{18}$OH    & $^2\Pi_{3/2} \hspace{0.2cm} J=5/2^--3/2^+$ & 119.97 &(8),(14)\\ 
$^{17}$OH    & $^2\Pi_{3/2} \hspace{0.2cm} J=5/2^+-3/2^-$ & 119.83 &  (12)\\ 
$^{17}$OH    & $^2\Pi_{3/2} \hspace{0.2cm} J=5/2^--3/2^+$ & 119.62 &  (12)\\ 
 
\enddata
\end{deluxetable}
\clearpage

\begin{deluxetable}{ccrr}
\tabletypesize{\scriptsize}
\tablecaption{Published Far--IR features detected with the LWS/FP
toward Sgr~B2(M) between 119.4-51.8~$\mu$m
\label{survey2}}
\tablewidth{0pt}
\tablehead{ 
\colhead{species} & \colhead{transition} & \colhead{$\lambda_{rest}$($\mu$m)} 
& \colhead{references}
}
\startdata
OH            & $^2\Pi_{3/2} \hspace{0.2cm} J=5/2^+-3/2^-$ &119.44& (2),(8),(15)\\ 
OH            & $^2\Pi_{3/2} \hspace{0.2cm} J=5/2^--3/2^+$ &119.23&(8),(15)\\ 
NH$_2$        & $2_{20}-1_{11} \hspace{0.2cm} J=5/2-3/2$ & 117.79 & (1)\\ 
NH$_2$        & $2_{20}-1_{11} \hspace{0.2cm} J=3/2-1/2$ & 117.38 &  (1) \\ 
NH$_2$        & $2_{20}-1_{11} \hspace{0.2cm} J=3/2-3/2$ & 117.07 &  (1)\\ 
H$_2$O        & $4_{14}-3_{03}$ & 113.54 &  (11)\\ 
HD            & $J=1-0$ & 112.07 &  (9)\\
H$_2$O        & $2_{21}-1_{10}$ & 108.07 &  (11)\\ 
H$_{2}^{18}$O & $2_{20}-1_{11}$ & 102.01 &  (11)\\ 
NH$_3$        & $(5,4)a-(4,4)s$ & 101.53 &  (7)\\ 
H$_2$O        & $2_{20}-1_{11}$  &  100.98&  (5), (11)\\
H$_3$O$^+$    & $2_0^--1_0^+$  & 100.87  &  (5)\\
H$_3$O$^+$    & $2_1^--1_1^+$ & 100.58 &  (5)\\
NH$_3$        & $(5,3)s-(4,3)a$ & 100.11 &  (7)\\
NH$_3$        & $(5,4)s-(4,4)a$ & 99.95  & (7)\\
OH            & $^2\Pi_{1/2} \hspace{0.2cm}5/2-3/2$ & 98.74&  (8)\\ 
H$_2$O        & $3_{22}-2_{11}$  & 89.99&  (11)\\ 
O$^{++}$      & $^3P_1-{^3P}_0$  & 88.36 &  (1)\\ 
OH            & $^2\Pi_{3/2} \hspace{0.2cm} J=7/2^--5/2^+$ & 84.60 &  (2),(8) \\
NH$_3$        & $(6,5)a-(5,5)s$ & 84.54 &  (7),(8)\\
OH            & $^2\Pi_{3/2} \hspace{0.2cm} J=7/2^+-5/2^-$ & 84.42&  (8)\\
NH$_3$        & $(6,5)s-(5,5)a$ & 83.43 &  (7)\\
OH            & $^2\Pi_{1/2}-{^2\Pi}_{3/2} \hspace{0.2cm}J=1/2^+-3/2^-$ & 79.18&  (8)\\
OH            & $^2\Pi_{1/2}-{^2\Pi}_{3/2}\hspace{0.2cm}J=1/2^--3/2^+$ & 79.12& (8)\\
H$_2$O        & $3_{21}-2_{12}$ & 75.38&  (11)\\
NH$_3$        & $(7,6)a-(6,6)s$ & 72.44 &  (7)\\
NH$_3$        & $(7,6)s-(6,6)a$ & 71.61 &  (7)\\
H$_2$O        & $3_{31}-2_{20}$ & 67.09&  (11)\\
H$_2$O        & $3_{30}-2_{21}$ & 66.44&  (11)\\
NH$_3$        & $(8,7)a-(7,7)s$ & 63.38 &  (7)\\
O$^0$         & $^3P_1-{^3P}_2$ &63.18 &  (6),(10)\\
NH$_3$        & $(8,7)s-(7,7)a$ & 62.73 &  (7)\\
NH$_3$        & $(9,8)a-(8,8)s$ & 56.34 &  (7)\\
OH            & $^2\Pi_{1/2}-{^2\Pi}_{3/2} \hspace{0.2cm}  J=3/2^--3/2^+$ & 53.35&  (8)\\
OH            & $^2\Pi_{1/2}-{^2\Pi}_{3/2} \hspace{0.2cm} J=3/2^+-3/2^-$ & 53.26  &  (8)\\
O$^{++}$      & $^3P_2-{^3P}_1$ & 51.82 &  (1)
\enddata
\tablerefs{
   (1) This paper; 
   (2) Cernicharo et al. 1997; 
   (3) Neufeld et al. 1997;
   (4) Cernicharo et al. 2000;
   (5) Goicoechea \& Cernicharo 2001;
   (6) Lis et al. 2001;
   (7) Ceccarelli et al. 2002;
   (8) Goicoechea \& Cernicharo 2002;
   (9) Polehampton et al. 2002; 
   (10) Vastel  et al. 2002;
   (11) Cernicharo et al., 2004 in prep.;
   (12) Polehampton et al., 2003 in prep.;
   (13) Stacey et al.  1987 (with KAO);
   (14) Lugten  et al. 1986 (with KAO);
   (15) Storey  et al. 1981 (with KAO);
   (16) Giesen  et al. 2001 (with KAO).
}
\end{deluxetable}
\clearpage

\end{document}